\documentclass[twocolumn,secnumarabic,amssymb, nobibnotes, aps, prd]{revtex4-1}

\usepackage{graphicx}
\usepackage{dcolumn}
\usepackage{bm}

\begin{document}

\title{Interplanar stiffness in defect-free monocrystalline graphite}%

\author{Koichi Kusakabe$^1$}
\author{Atsuki Wake$^2$}
\author{Akira Nagakubo$^2$}
\author{Kensuke Murashima$^3$}
\author{Mutsuaki Murakami$^3$}
\author{Kanta Adachi$^4$}
\author{Hirotsugu Ogi$^2$}
\email[]{ogi@prec.eng.osaka-u.ac.jp}
\affiliation{$^1$Graduate School of Engineering Science, Osaka University, Toyonaka, Osaka 560-8531, Japan}
\affiliation{$^2$Graduate School of Engineering, Osaka University, Suita, Osaka 565-0871, Japan}
\affiliation{$^3$Material Solutions New Research Engine E \& I Materials Group, Kaneka Co., Settsu, Osaka 566-0072, Japan}
\affiliation{$^4$Faculty of Science and Engineering, Iwate University,  Morioka, Iwate 020-8551, Japan}

\date{\today}%

\begin{abstract}
The interplanar bond strength in graphite has been identified to be very low owing to the contribution of the van der Waals interaction. However, in this study, we use microscopic picosecond ultrasound to demonstrate that the elastic constant, $C_{33}$, along the $c$ axis of defect-free monocrystalline graphite exceeds 45 GPa, which is higher than reported values by 20\%. Existing theories fail to reproduce this strongly correlated interplanar system, and our results, thus, indicate the necessity for improvement. 
Since the LDA+U+RPA method, including both random phase approximation correlation and short-range correlation in $p$ Wannier orbitals, shows better agreement with the observation than LDA or even than ACFDT-RPA, the experimental results indicate non-negligible electron correlation effects with respect to both the short-range and long-range interactions. 
\end{abstract}

\maketitle
The interplanar interaction between graphene sheets remains a central issue in condensed matter physics because the participation of long-distance van der Waals interactions makes its theoretical description a labyrinth problem. A direct characteristic of an interplanar interaction is $C_{33}$ - the elastic constant along the $c$ axis of graphite - because it reflects the interlayer bond strength. Lechner et al. \cite{Lechner} demonstrated that $C_{33}$ of graphite estimated by density functional theory (DFT) can range between 1.9 and 71.4 GPa depending on the Hamiltonian basis set used. $C_{33}$ of graphite has thus been adopted to validate proposed theoretical approaches, and its accurate measurement is critical to thoroughly understanding van der Waals interactions. 

As presented in Table I, previous experimental studies reported $C_{33}$ values between 36.5 and 38.7 GPa \cite{Blakslee, Wada, Nicklow, Bosak}, and recent DFT studies were conducted to yield the interplanar stiffness \cite{RPA, Lebedeva, Michel}. However, the specimens used in the studies were highly oriented pyrolytic graphite (HOPG), not defect-free monocrystalline graphite. Due to the hexagonal symmetry about the $c$ axis of graphite, an HOPG specimen was apparently regarded as a single crystal \cite{Blakslee}. However, grain (domain) boundaries usually deteriorate the bond strength of the material, significantly decreasing the macroscopic elastic constants. Young's moduli of nanocrystalline CaF$_2$, Pd, and Mg decrease by 66\%, 35\%, and 13\%, respectively, from those of corresponding monocrystals \cite{Korn} predicted by the Hill averaging method \cite{Hill}. In addition, Young's moduli of nanocrystalline Pd and Cu\cite{Nieman}, Fe\cite{Bonetti}, and Al\cite{Zhang} are reported to decrease by approximately 50\%, 70\%, 40\%, and 40\%, respectively. 

A number of other studies have been conducted on softened materials by grain boundaries. Thus, the macroscopic elastic constant of HOPG should be substantially smaller than that of monocrystalline graphite. Jansen and Freeman \cite{Jansen} performed an all-electron total-energy local-density calculation and determined the elastic constants of graphite, which were significantly different from the previous experiment. Consequently, they emphasized the need for experiments with a defect-free monocrystalline sample. Bosak et al. \cite{Bosak} used focused inelastic X-ray scattering to measure an area of 250$\times$60 $\mu$m$^2$, producing the largest $C_{33}$ value among previous studies: 38.7 GPa. However, this size appeared too large to express the stiffness of a single grain, and experimental reports of $C_{33}$ in defect-free monocrystalline graphite were still lacking.

\begin{table*}
\caption{Measured and calculated $C_{33}$ in the present study and previous studies. Simulation methods are described in the main text.}
\begin{ruledtabular}
\begin{tabular}{cclccl}
 &  & $C_{33}$ (GPa) & Thickness ($\mu$m) & Domain size ($\mu$m) & Methods\\
\hline
 & present ($G_{2800}$) & 40.1 $\pm$0.9 & $\sim$1.5  & 1.3 & \\
 & present ($G_{3150}$) & 46.1 $\pm$4.4 & $\sim$1.2  & 6.3 & Picosecond ultrasonics\\
 & present ($G_{3200}$) & 48.4 $\pm$5.3 & $\sim$1.3  & 8.3 &  \\
exp. & Ref. \cite{Blakslee} & 36.5 & $\sim$10,000  & - & Ultrasonic pulse echo\\
 & Ref. \cite{Wada} & 36.6$^a$ & $\sim$50  & - & X-ray diffraction\\
 & Ref. \cite{Nicklow} & 37.1 & -  & - & Inelastic neutron scattering with lattice dynamics\\
 & Ref. \cite{Bosak} & 38.7 & $\sim$100  & - &Inelastic X-ray scattering\\
\vspace{-2mm}
\\
 & present & 37 $\pm$8 & -  & - &LDA+U ($U= $2.1[eV])\\
 & present & 45 $\pm$8 & -  & - &LDA+U+RPA ($U= $2.1[eV])\\
 & present & 38 & -  &  - &ACFDT-RPA\\
 & present & 59 $\pm$5 & -  &  - &ACFDT-RPA+U ($U= $2.1[eV]) \\
 calc.  & Ref.\cite{RPA} & 36 & -  &  - &ACFDT-RPA\\
 & Ref. \cite{Lebedeva} & 33.3 & -  &  - &vdW-DF2\\
 & Ref. \cite{Michel} & 38.7 & -  &  - &Born's long-wave method\\
 & Ref. \cite{Mounet} & 45 & -  &  - &GGA with measured lattice constants\\
 & Ref. \cite{Jansen} & 56.9 $\pm$9 & -  &  - &All-electron total-energy LDA\\
\end{tabular}
\end{ruledtabular}
\begin{flushleft}
\scriptsize{
}
\end{flushleft}
\end{table*}

In this study, we measured $C_{33}$ of highly pure defect-free monocrystalline graphite using a microscopic picosecond-ultrasonic method. Our specimens were multilayer graphene sheets that were synthesized by heating $\sim$3-$\mu$m-thick polyimide films at temperatures up to 2,800, 3,150, and 3,200 $^\circ$C under in-plane tension \cite{Kaneka1, Kaneka2, Kaneka3}; we refer to these sheets as $G_{2800}$, $G_{3150}$, and $G_{3200}$, respectively. This novel synthesis method allowed us to develop approximately 1.5-$\mu$m-thick highly oriented defect-free graphite specimens with a domain (grain) size up to approximately 20 $\mu$m, as illustrated in Fig. 1. As demonstrated in Supplementary Material A, our microscopic picosecond ultrasonic method can measure the longitudinal wave velocity along the thickness direction in a localized area of approximately 1 $\mu$m in diameter, which is smaller than the domain size of the $G_{3150}$ and $G_{3200}$ specimens (Table I), resulting in $C_{33}$ of defect-free monocrystalline graphite.

\begin{figure}[tp]
\begin{center}
\includegraphics[width=90mm]{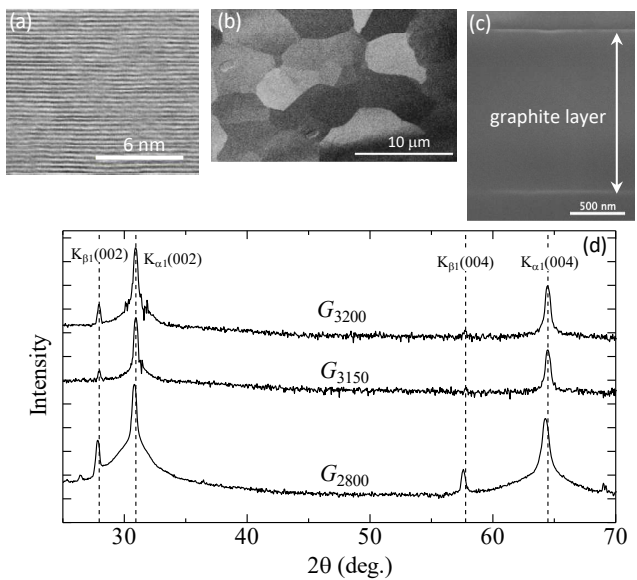}
\end{center}
\caption{(a) Cross-section transmission electron microscopy image for specimen $G_{2800}$; (b) in-plane scanning electron microscopy image for specimen $G_{3200}$; (c) cross-section scanning electron microscopy image for specimen $G_{3200}$; (d) X-ray diffraction spectra of three graphite specimens (Co target).}
\label{Fig. 1}
\end{figure}

\begin{figure}[tp]
\begin{center}
\includegraphics[width=80mm]{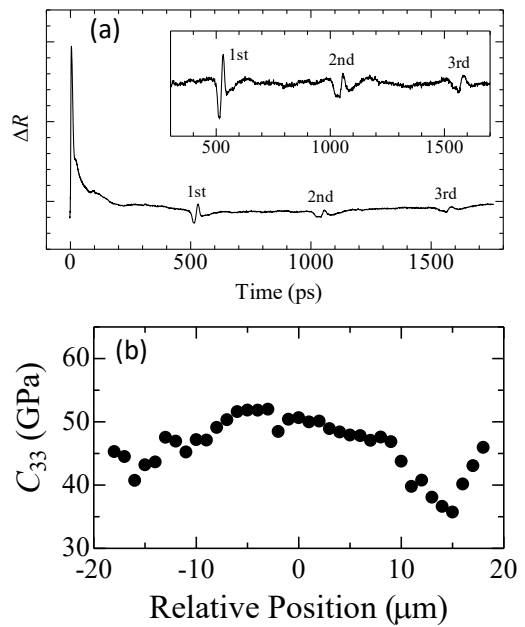}
\end{center}
\caption{(a) Reflectivity change measured using the microscopic picosecond ultrasonic method for $G_{3150}$; (b) one-dimensional elastic-constant distribution on the specimen.}
\label{Fig. 2}
\end{figure}

The optics developed are described in Supplementary Figure SA1. We used a titanium/sapphire pulse laser with a wavelength of 800 nm. The light pulse was split into pump and probe light pulses, and the wavelength of the probe light was converted to 400 nm. Both pulses were perpendicularly focused on the specimen surface via an objective lens. Due to the high absorption coefficient of graphite for 800-nm light, a longitudinal wave was efficiently generated without requiring any additional coating material.  

Figure 2(a) illustrates the typical reflectivity change, and the inset presents the baseline subtracted reflectivity. We were able to clearly observe the pulse-echo signals of the longitudinal wave propagating along the thickness direction, from which we determined the round-trip time and elastic constant $C_{33}$ using the mass density, 2,260 kg/m$^3$, and the specimen thickness, which is a key parameter for determining the elastic constant. After the picosecond ultrasonic measurement, we lifted a small slice of the specimen near the surface using a focused ion beam processing technique, which is widely used to prepare a specimen for transmission electron microscopy for cross-sectional observation. We observed its cross-section by electron microscopy, as illustrated in Fig. 1(c). Figure 2(b) presents a line scan of the elastic constant. The $C_{33}$ value is generally within 45-50 GPa; however, it is occasionally significantly smaller. We consider these softened regions to correspond to measurements near the domain boundary, and thus exclude them in the stiffness determination.  The $C_{33}$ values determined in this study are presented in Table I. They are significantly higher than in previous reports, which we attribute to reduced stiffness due to defects in previous studies. The stiffness of the $G_{2800}$ specimen is lower than those of $G_{3150}$ and $G_{3200}$ specimens, because its domain size is nearly the same as the measurement spot involving the domain-boundary affected region.  

Recent DFT calculations \cite{RPA, Lebedeva, Michel} have failed to yield our determined $C_{33}$ value ($>$ 40 GPa). For example, we used the adiabatic-connection fluctuation-dissipation-theorem with random-phase approximation (ACFDT-RPA) calculation following Leb\'egue et al. \cite{RPA} and confirmed that the deduced $C_{33}$ value could not exceed 39 GPa, as illustrated in Table I. One notable DFT result is the estimation of interplanar stiffness using the second derivatives of the energy-strain relationship using generalized gradient approximation Perdew-Burke-Ernzerhof (GGA-PBE) at experimental lattice constants \cite{Mounet}, which yielded a $C_{33}$ value of 45 GPa. A DFT calculation using generalized gradient approximation (GGA) for graphite is inappropriate because it highly overestimates the lattice constant along the $c$ axis, resulting in an impossibly small elastic constant. The results of Mounet-Marzari, however, indicate that the DFT calculation using GGA is suitable with fixed experimental lattice constants with respect to the second derivative of each energy contribution. This applies to other materials as well. For example, DFT calculations using GGA for the interplanar stiffness $C_{33}$ of GaN and $\beta$-Ga$_2$O$_3$ produce significantly smaller values than experimental values because they overestimate the lattice constant along the $c$ axis \cite{AdachiJAP1, AdachiJAP2}. However, their $C_{33}$ values demonstrate good agreement with experimental values by calculating the second derivatives at the experimental lattice constants. The difference between the calculated and experimental values is 9.6\% for GaN and 12\% for $\beta$-Ga$_2$O$_3$ with GGA-PBE; however, this difference becomes 5.2\% and 0.3\%, respectively, by using the experimental lattice constants, as illustrated in Supplementary Material A. Thus, our experimental results for $C_{33}$, which are significantly higher than those of previous experiments, are supported by these calculations as well as the fact that defect-free materials exhibit higher (ideal) stiffness. 

We propose a perturbation method coupled with the estimation of correlation effects in both long- and short-range schemes. This method improves the full DFT local density approximation (DFT-LDA) calculation for both $C_{33}$ and the lattice constant of graphite. In Supplementary Material B, we introduce a scheme for the use of the many-body perturbation approach starting from the LDA+U mean-field calculation. Using the mean-field wave function given by the LDA+U method, we substitute LDA correlation with ACFDT-RPA correlation. This treatment is justified when all Feynman diagrams remain unchanged except for reduction of 
the scattering amplitudes in the $\lambda$ integration of the ACFDT-RPA calculation. The reduction is caused by shift in the mean-field contribution counted with respect to the on-site-U term. We can check the validity by finding similarity in the single-particle excitation (the band structure of graphite given by LDA+U) to the original LDA (or GGA) spectrum. We call this approach LDA+U+RPA. 

We can also introduce residual short-range correlation into the ACFDT-RPA calculation. Assuming that calculation of ACFDT-RPA using Kohn-Sham local density approximation (LDA) or GGA wave functions is an estimation with a self-consistently converged charge density, the Hubbard correlation energy estimated by the same wave function can be added as an additional $\lambda$ integration. Therefore,  the ACFDT-RPA+U energy is approximated by the ACFDT-RPA result added by the Hubbard correlation estimated in an LDA+U calculation, in which mean-field approximation for the Hubbard correlator is used. LDA+U+RPA and ACFDT-RPA+U are mutually related with a closed integration path in the space of MR-DFT models. 

\begin{figure}[tp]
\begin{center}
\includegraphics[width=65mm,angle=270]{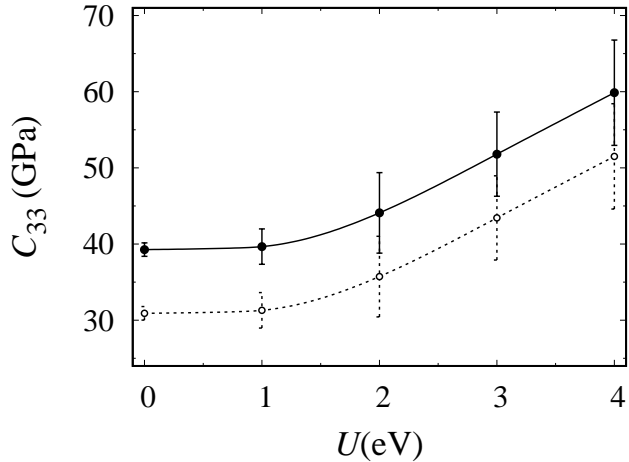}
\end{center}
\caption{Calculated $C_{33}$ by LDA+U with and without RPA correction. The horizontal axis denotes the strength of $U$[eV]. Closed (open) circles are given LDA+U with (without) RPA correction. The vertical bar at each point denotes a range of the estimated $C_{33}$ 
by several k-mesh points from $14\times14\times6$ to $44\times44\times22$ points.  Constrained RPA (cRPA) estimation results in an estimated value of $U\sim 2.1\pm0.5$ [eV] for a Wannier orbital with $p_z$ symmetry.}
\label{Fig. 3}
\end{figure}

In this study, we used a scheme from multi-reference density functional theory \cite{Kusakabe, Kusakabe2, Kusakabe-Maruyama} to allow for the simultaneous determination of both RPA-screened $U$ and the ACFDT-RPA correlation. The theoretical background is provided in Supplementary Material B. As an approximation, we performed calculations of the LDA+U scheme on a Wannier orbital with the double-counting correction \cite{Cococcioni}. The Wannier orbitals with $p$ symmetry on each carbon were determined for each deformed lattice structure on the Born-Oppenheimer energy surface. This self-consistent determination of the Wannier function and the resulting LDA+U mean-field ground state produced rather large corrections of the value of $C_{33}$ (Fig.~\ref{Fig. 3}). 

Using the LDA+U+RPA scheme, an approximated value of $C_{33}$ is produced. There remains residual dependence on the Brillouin zone integration with respect to the $k$ point mesh. A finer mesh for the LDA+U calculation results in a larger $C_{33}$ value. Using ACFDT-RPA correction for $C_{33}$ as noted above and adopting the value of $U\simeq 2.1 \pm 0.5$ eV by a constrained RPA (cRPA) evaluation of $U$, we have $C_{33}\simeq 45\pm 8$ GPa. Upon adding the Hubbard correction using the self-consistent calculation by LDA for the ACFDT-RPA calculation, we obtain an ever larger $C_{33}$ of approximately $59 \pm 5$ GPa. This ACFDT-RPA+U approximation, however, may contain an overestimation error because the wave function adopted for the +U correction is not used for the mean-field ground state energy containing kinetic and electron-ion energy contributions.

In the above estimation, the value of $U$ in cRPA is assumed to be independent of the $c$ axis lattice parameter, $c$. In fact, there exists a linearly dependent shift in $U$ around the equilibrium structure, whereby a smaller $c$ leads to a larger $U$ in cRPA. Then, the total energy curve is shifted by a linear contribution from the Hubbard term and the double-counting correction term. Although this effect can lead to an increased evaluated $c$, $C_{33}$, as the second-order derivative of the total energy $E_{\rm tot}$ with respect to $c$, is not affected by the correction linear to $U$. The correction stems from  $\lambda$ integration with respect to the difference between the Hubbard terms, that is, the Hubbard interaction and the double-counting correction, and their mean-field approximations. By applying a constant $U$ approximation, however, we obtain a reasonable result for the estimated volume within an accuracy of several percent, as described in Supplementary Material B. 

In summary, we use microscopic picosecond ultrasound measurement to demonstrate that defect-free monocrystalline graphite exhibits a $C_{33}$ value above 45 GPa, which exceeds the value estimated by ACFDT-RPA. Considering the short-range correlation effect, the theoretical estimation of $C_{33}$ can produce a value larger than 45 GPa, as exemplified by our proposed LDA+U+RPA or ACFDT-RPA+U methods.

\subsection*{ACKNOWLEDGEMENTS}
This study is supported by JSPS KAKENHI Grant Nos. JP19H00862 and JP18K03456. 
The calculations were done in the computer centers of Kyushu University and ISSP, University of Tokyo. 

\subsection*{Contributions}
K. Kusakabe performed theoretical estimation of $C_{33}$ with MR-DFT and wrote the paper.   A. Wake and A. Nagakubo performed the picosecond ultrasonics measurements.  K. Murashima and M. Murakami synthesized the defect-free graphite specimens.  K. Adachi performed the ab-initio calculation of elastic constants of GaN and $\beta$-Ga$_2$O$_3$.  H. Ogi produced this study, performed the picosecond ultrasonics measurements, analyzed specimens, and wrote the paper. 

\clearpage

\end{document}


\title{Supplementary Material A of ``Interplanar stiffness in defect-free monocrystalline graphite''}%

\author{Koichi Kusakabe$^1$}
\author{Atsuki Wake$^2$}
\author{Akira Nagakubo$^2$}
\author{Kensuke Murashima$^3$}
\author{Mutsuaki Murakami$^3$}
\author{Kanta Adachi$^4$}
\author{Hirotsugu Ogi$^2$}
\email[]{ogi@prec.eng.osaka-u.ac.jp}
\affiliation{$^1$Graduate School of Engineering Science, Osaka University, Toyonaka, Osaka 560-8531, Japan}
\affiliation{$^2$Graduate School of Engineering, Osaka University, Suita, Osaka 565-0871, Japan}
\affiliation{$^3$Material Solutions New Research Engine E \& I Materials Group, Kaneka Co., Settsu, Osaka 566-0072, Japan}
\affiliation{$^4$Faculty of Science and Engineering, Iwate university, Morioka, Iwate 020-8551, Japan}

\date{\today}%

\begin{abstract}
We present details of the microscopic picosecond ultrasound and elastic-constant calculations for GaN and $\beta$-Ga$_2$O$_3$ with lattice parameters fixed at their experimental values. 
\end{abstract}

\maketitle

\section{Microscopic Picosecond Ultrasounds}

\begin{table*}
\captionsetup{labelformat=empty,labelsep=none}
\caption{TABLE SA1: Elastic constants (GPA) of GaN and $\beta$-Ga$_2$O$_3$.  DFT calculations are performed with PAW-PBE using the VASP package.}
\begin{ruledtabular}
\begin{tabular}{ccccccccccccccccc}
 &  & $C_{11}$ & $C_{22}$ & $C_{33}$ & $C_{44}$ & $C_{55}$ & $C_{66}$ & $C_{12}$ & $C_{13}$ & $C_{23}$ & $C_{15}$ & $C_{25}$ & $C_{35}$ & $C_{45}$ & \\
\hline
        & exp.  & 359.7 & - & 391.8 & 99.6 & - & 114.9 & - & 104.6 & - & - & - & - & - & \\
GaN & calc. & 328.0 & - & 354.0 & 86.0 & - & 100.0 & - & 95.0 & - & - & - & - & - & \\
         & calc. & 380.5 & - & 412.3 & 90.6 & - & 104.8 & - & 133.3 & - & - & - & - & - & with fixed lattice constants\\
\vspace{-2mm}
\\
 & exp.  & 242.8 & 343.8 & 347.4 & 47.8 & 88.6 & 104.0 & 128.3 & 160.0 & 70.9 & -1.62 & 0.36 & 0.97 & 5.59\\
$\beta$-Ga$_2$O$_3$ & calc.  & 204.0 & 323.7 & 304.8 & 44.9 & 72.9 & 92.6 & 115.9 & 138.5 & 78.4 & -1.30 & 2.11 & 16.74 & 7.78\\
 & calc.  & 220.9 & 384.2 & 346.5 & 59.3 & 76.6 & 100.4 & 133.3 & 188.2 & 119.1 & -3.95 & 2.62 & 15.80 & 13.78 & with fixed lattice constants\\
\end{tabular}
\end{ruledtabular}
\begin{flushleft}
\scriptsize{
}
\end{flushleft}
\end{table*}

Figure SA1 illustrates the optics of the picosecond ultrasonics developed in this study. We used a titanium-sapphire pulse laser (Chameleon, Coherent Inc.) with a center wavelength of 800 nm, pulse duration time of 200 fs, and pulse repetition rate of 80 MHz. The light pulse was split into pump and probe light pulses by a polarizing beam splitter (PBS) after a $\lambda/2$ half plate. After the PBS, the 800-nm light pulse, which was used as the pump light, entered the corner reflector to create a delay line and then entered the acousto-optic modulator, through which the power was modulated at 100 kHz.  After the power modulation, the pump light pulse was reflected by the dichroic mirror (DM) and focused on the specimen surface.  

The wavelength of the light pulse reflected by the PBS was converted to 400 nm by a second-harmonic generator (SHG). Part of the 400-nm light pulse was reflected by a beam splitter (BS) and detected at the reference channel of the balance detector. The major 400-nm light pulse, which was used as the probe light pulse, travelled through the DM and was focused on the specimen surface via an objective lens with a magnification between 50 and 150.  The probe light reflected at the specimen surface, which involved information of the elastic wave generated by the pump light pulse, entered the signal channel in the balanced detector. The balanced output was then connected to a lock-in amplifier to extract the modulated components. By changing the delay line, we were able to obtain the time-resolved reflectivity change, as in Fig. 2(a), which indicates the pulse-echo signals of the longitudinal wave propagating along the $c$ axis. Figure SA2 presents an irradiation image, in which a measurement spot of approximately 1.5 $\mu$m in diameter is indicated.

\begin{figure}[tp]
\begin{center}
\includegraphics[width=85mm]{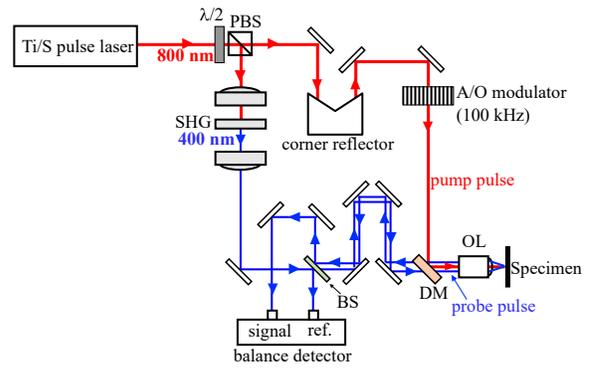}
\end{center}
\captionsetup{labelformat=empty,labelsep=none}
\caption{Fig. SA1.  Microscopic picosecond ultrasound optics.}\label{Fig. SA1}
\end{figure}

\begin{figure}[]
\begin{center}
\includegraphics[width=50mm]{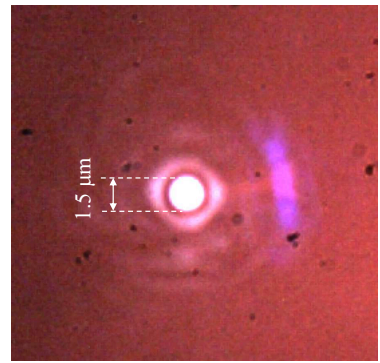}
\end{center}
\captionsetup{labelformat=empty,labelsep=none}
\caption{Fig. SA2.  Irradiation of gauged specimen with microscopic picosecond ultrasound.}
\label{Fig. SA2}
\end{figure}

The deposition of a metallic thin film on the specimen surface is usually required as the acoustic-wave transducer. However, additional coating was not required for our graphite specimens; we directly excited and detected the longitudinal wave, thus simplifying the travel time analysis. 

\section{Density Functional Theory Calculation for Elastic Constants of gallium nitride and $\beta$-gallium oxide}
Table SA1 presents the elastic constants of GaN and $\beta$-Ga$_2$O$_3$ calculated using the Vienna Ab initio Simulation Package (VASP). GGA-PBE potentials were used for the exchange correlation. The plane-wave cutoff energy was 1,300 eV, and k mesh points of $\times$10$\times$10$\times$10 were adopted. As GaN and $\beta$-Ga$_2$O$_3$ possess five and 13 independent elastic constants, respectively, we applied five and 13 different deformation modes (elongation, shearing, breathing, etc.), respectively, to their unit cells. In each deformation mode, we applied corresponding strain up to 1\%, and relaxed the ions inside the cell. We thus obtained the relationships between the total energy and the applied strain. By fitting the harmonic function to these relationships, we determined the effective stiffnesses, through which all of the independent elastic constants were obtained. Because the DFT calculation overestimated the lattice constants, the resulting elastic constants were generally larger than the experimental values. However, when we derived the elastic constants at the experimental lattice constants, they demonstrated better agreement with the experimental results.


\title{Supplementary Material B of ``Interplanar stiffness in defect-free monocrystal graphite''}%

\author{Koichi Kusakabe$^1$}
\author{Atsuki Wake$^2$}
\author{Akira Nagakubo$^2$}
\author{Kensuke Murashima$^3$}
\author{Mutsuaki Murakami$^3$}
\author{Kanta Adachi$^4$}
\author{Hirotsugu Ogi$^2$}
\email[]{ogi@prec.eng.osaka-u.ac.jp}
\affiliation{$^1$Graduate School of Engineering Science, Osaka University, Toyonaka, Osaka 560-8531, Japan}
\affiliation{$^2$Graduate School of Engineering, Osaka University, Suita, Osaka 565-0871, Japan}
\affiliation{$^3$Material Solutions New Research Engine E \& I Materials Group, Kaneka Co., Settsu, Osaka 566-0072, Japan}
\affiliation{$^4$Faculty of Science and Engineering, Iwate university, Morioka, Iwate 020-8551, Japan}

\date{\today}%

\begin{abstract}
We present the theoretical background and formulation for the LDA+U+RPA and 
ACFDT-RPA+U methods. 
\end{abstract}

\maketitle

\section{LDA+U approach derived from multi-reference density functional theory}

In this section, we review a theoretical method defining a model Hamiltonian with the Hubbard term 
within the framework of multi-reference density functional theory (MR-DFT) \cite{Kusakabe,Kusakabe2,Kusakabe-Maruyama}. 
A model exchange-correlation energy density functional $E_{\rm xc}[n]$ 
in addition to the Hartree energy functional 
is effective for determining the orbital when the electron density distribution $n({\bf r})$ 
is given by a trial multi-reference state along the simulation steps of MR-DFT. 
In local density approximation (LDA) 
or generalized gradient approximation (GGA), 
an exchange-correlation energy functional can be selected for the definition of $E_{\rm xc}[n]$. 
Adopting MR-DFT, we can introduce a quantum-fluctuation term explicitly in the energy functional. 
When the on-site density-density correlation is considered, the Hubbard interaction appears. 
Therefore, an LDA+U Hamiltonian is obtained. In our formalism, the mean-field description of 
LDA+U has an explicit difference in the form of the state vector from former 
theoretical approaches \cite{Anisimov,Anisimov2,Solovyev}. 
Specifically, the LDA+U ground state $|\Phi_{\rm MF}\rangle$ is defined to be a multi-reference state. 

\subsection{Representation of correlated electron state} 

We start from the electron Hamiltonian in an external potential, which is represented by 
$\hat{V}_{\rm ext}$. 
\begin{equation}
\hat{\cal H}_{\rm Coulomb} = \hat{T} + \hat{V}_{\rm ee} + \hat{V}_{\rm ext}. 
\end{equation}
Here, the kinetic energy operator $\hat{T}$ is assumed to be a non-relativistic form. 
\begin{equation}
\hat{T}=- \frac{\hbar^2}{2m} \int d^3r {\rm lim}_{{\bf r}'\rightarrow {\bf r}}
\hat{\psi}_{\sigma}^\dagger ({\bf r}) \Delta_{{\bf r}'}
\hat{\psi}_{\sigma} ({\bf r}'), 
\end{equation}
where the electron field operators $\hat{\psi}_{\sigma}^\dagger ({\bf r})$ (and 
$\hat{\psi}_{\sigma} ({\bf r})$) are expanded by a complete set of basis functions 
of $L^2({\mathbb R}^3)$ and their electron creation (annihilation) operators. 
If necessary, $\hat{T}$ can be replaced by the Dirac operator.  
The electron-electron interaction $\hat{V}_{\rm ee}$ is assumed to be 
an interaction operator in the form of a Coulomb interaction. 
\begin{equation}
\hat{V}_{\rm ee} = \frac{e^2}{2} \iint d^3r d^3r' \frac{1}{|{\bf r}-{\bf r}'|} 
: \hat{n}({\bf r}')\hat{n}({\bf r}') :.
\end{equation}
Here, 
the density operator is given by the electron field operator,
\begin{equation}
\hat{n}({\bf r}) \equiv \sum_\sigma \hat{\psi}_{\sigma}^\dagger ({\bf r}) \hat{\psi}_{\sigma} ({\bf r}) ,
\end{equation}
and the symbol 
\[:\hat{\psi}_{\sigma}^\dagger ({\bf r}) \hat{\psi}_{\sigma} ({\bf r}) \hat{\psi}_{\sigma'}^\dagger ({\bf r}') \hat{\psi}_{\sigma'} ({\bf r}'):\equiv\hat{\psi}_{\sigma}^\dagger ({\bf r}) \hat{\psi}_{\sigma'}^\dagger ({\bf r}') \hat{\psi}_{\sigma'} ({\bf r}')\hat{\psi}_{\sigma} ({\bf r}),\]
represents the normal ordering operation. 

The Schr\"{o}dinger equation for the stationary ground state is given by 
\begin{equation}
\hat{\cal H}_{\rm Coulomb} |\Psi_0\rangle = E_0 |\Psi_0\rangle .
\end{equation}
Here, we assume that the lowest stable state $|\Psi_0\rangle$ exists and it has ground-state energy $E_0$. 
The state vector $|\Psi_0\rangle$ is expanded in a series of state vectors, and 
its representation is a multi-Slater form of the wave function. 

To obtain the explicit form of the representation, we use an effective potential and 
its eigenfunctions to create an expanding basis set. In condensed matter physics, 
there exist multiple recipes for creating the potential, $v_{\rm eff}({\bf r})$. 
Among these recipes, the density functional theory (DFT) provides us a special merit. 
When $|\Psi_0\rangle$ is known to exist, we obtain the charge density $n_{\Psi_0}({\bf r})$ as 
the expectation value of $n_{\Psi_0}({\bf r})\equiv \langle \Psi_0|\hat{n}({\bf r})|\Psi_0\rangle$. 
The DFT effective potential is derived by adopting an exchange-correlation energy functional. 
Usually, this effective potential problem produces a set of single-particle wave functions 
$\{\psi_{k}({\bf r})\}$ expanding $L^2({\mathbb R}^3)$ 
and the single-particle energy spectrum $\varepsilon_{k}$, which possesses a band structure. 
Therefore, we can use the concepts of core levels, valence levels, conduction levels, and 
high-energy bands. 
The band structure allows us to define a useful projection operator $P_A$ onto correlated states having 
a common core state. The core state is composed of filled core levels without core holes. 

For an $N$ electron system with electron number $N$, 
the construction method of the complete set of $N$ electron wave functions is given 
by introducing $\{\psi_{k}({\bf r})\}$, 
spin wave functions $\xi(\sigma)=\alpha(\sigma), \beta(\sigma)$ 
($\sigma=\uparrow, \downarrow$), 
and the anti-symmetrization method. 
In mean-field approximation, an independent particle picture is provided. 
The approximated mean-field counterpart of $|\Psi_0\rangle$ becomes 
a single Slater determinant composed of $\{\psi_{k}({\bf r})\xi(\sigma)\}$. 
Following spectrum $\varepsilon_{k}$, 
the occupation number of electrons in the band structure is determined.
Namely, the single-particle description determines the Fermi energy $E_F$ 
as the highest energy levels with finite occupation. Similarly, in the full state vector $|\Psi_0\rangle$, 
occupation of levels far below $E_F$ is full filling in an effective manner. 

Among the vectors in the series expansion of $|\Psi_0\rangle$, 
we often observe contributions by a state with filled core levels and 
fully unoccupied high-energy bands for electrons. 
Thus, we can introduce a projection operator $P_A$, which projects any state 
with a hole in this core state or an electron in high-energy bands. 
Therefore, we obtain a representation of $|\Psi_0\rangle$ using $P_A$ and $P_B\equiv \hat{1}-P_A$ as follows: 
\begin{equation}
|\Psi_0\rangle = P_A |\Psi_0\rangle + P_B |\Psi_0\rangle.
\end{equation}
In other words, by selecting a proper recipe, 
it is possible to determine a suitable A state by choosing $P_A$ so that 
\begin{equation}
P_A |\Psi_0\rangle \neq 0.
\end{equation} 
In this manner, we can separate the A state, $P_A|\Psi_0\rangle$, from the high-energy excited B state, 
$(\hat{1}-P_A)|\Psi_0\rangle$. 

Here, we implicitly assume that in $P_A|\Psi_0\rangle$, 
there appears a core state vector $| \Psi_{0,{\rm core}}\rangle$ expressed by a single Slater determinant composed of only filled orbitals. 
The next direct product form, which is implicitly introduced above, is as follows: 
\begin{equation}
P_A |\Psi_0 \rangle = |\Psi_{0,{\rm corr}} \rangle \otimes | \Psi_{0,{\rm core}}\rangle 
\end{equation}
In general, the electron correlation effect is represented by the multi-reference correlated state $|\Psi_{0,{\rm corr}}\rangle$, whose expanding orbitals have orbital energies close to the Fermi energy. 

\subsection{MR-DFT functional and effective Hamiltonian} 

To derive an effective many-body electron model, we can start from MR-DFT when the charge-fluctuation term is 
introduced \cite{Kusakabe,Kusakabe2,Kusakabe-Maruyama}.
We then have a model energy functional as follows:
\begin{eqnarray}
\lefteqn{G_{\rm MR-DFT}[\Psi] } \nonumber \\
&=& \langle \Psi | \hat{T} | \Psi \rangle 
+E_{\rm Hartree} [n_\Psi] +E_{\rm xc}[n_\Psi] \nonumber \\
&+& \langle \Psi | \hat{V}_{\rm ee} - \hat{V}_{BB} | \Psi \rangle 
-E_{\rm dc}^{(1)}[\Psi]  
+E_{\rm ext}[n_\Psi]. 
\label{MR-DFT_functional}
\end{eqnarray}
This is a functional of the wave function of the state vector $|\Psi\rangle$. 
We can then appropriately select the representation 
of the wave function either in a real space or a reciprocal space (momentum space). 
If necessary, the Wannier representation in the second quantization form can be introduced 
after having the Bloch representation of the crystal structure of a material. A one-particle wave equation is introduced below. 

In the definition of $G_{\rm MR-DFT}[\Psi]$, the charge density of $|\Psi\rangle$ is defined as
\begin{equation}
n_\Psi ({\bf r})= \langle \Psi | \hat{n}({\bf r}) | \Psi \rangle. 
\end{equation}
Using the density, the Hartree energy functional is given as 
\begin{equation}
E_{\rm Hartree} [n_\Psi] 
= \frac{1}{2} \iint d^3rd^3r' \frac{e^2}{\left|{\bf r}-{\bf r}' \right|}n_\Psi ({\bf r})n_\Psi ({\bf r}'). 
\end{equation}

The exchange-correlation energy density functional $E_{\rm xc}[n_\Psi]$ is a negative definite 
functional of $n_\Psi({\bf r})$.
For this definition, because we have the spin-paramagnetic orbital-diamagnetic 
ground state of graphite, we omit the spin density dependence and use the functional.
\begin{equation}
E_{\rm xc} [n_\Psi] = \int d^3r f(n_\Psi({\bf r}), \nabla n_\Psi({\bf r})).
\end{equation}
In LDA, using the exchange-correlation energy density $\varepsilon_{\rm xc} (n)$, 
\[f(n_\Psi({\bf r}), \nabla n_\Psi({\bf r}))=\varepsilon_{\rm xc} (n_\Psi ({\bf r})) n_\Psi ({\bf r}).\] 
For GGA, we use the Perdew-Burke-Ernzerhof (PBE) functional in this study. 
The external potential contribution $E_{\rm ext}[n_\Psi]$ denotes the potential energy from 
the external potential $v_{\rm ext}({\bf r})$. 
\begin{equation}
E_{\rm ext}[n_\Psi] = \int d^3r v_{\rm ext}({\bf r}) n_{\Psi}({\bf r}). 
\end{equation}

The fourth and fifth terms of (\ref{MR-DFT_functional}), 
\begin{equation}
E_{\rm fluc}[\Psi] = \langle \Psi | \hat{V}_{\rm ee} - \hat{V}_{BB} | \Psi \rangle-E_{\rm dc}^{(1)}[\Psi], 
\end{equation}
represent a quantum fluctuation term that is active in 
the correlated state of $|\Psi_{0,{\rm corr}}\rangle$.
The partial interaction term $\hat{V}_{BB}$ represents inter-quasi-particle Coulomb 
scattering processes, in which all particles involved in the process are quasi-holes in the core levels or 
quasi-electrons in the uncorrelated higher orbitals where the occupation is zero in $P_A |\Psi\rangle$. 
Therefore, for any state $|\Psi\rangle$, $\hat{V}_{BB} P_A |\Psi\rangle = 0$. 
The definitions of $\hat{V}_{BB}$ and $P_A$ are mutually correlated. 
\begin{eqnarray}
P_A \hat{V}_{BB} P_A &=& P_B \hat{V}_{BB} P_A = P_A \hat{V}_{BB} P_B = 0, \\
P_B \hat{V}_{BB} P_B &=& \hat{V}_{BB}.
\end{eqnarray}

The separation of the Coulomb interaction operator is defined as follows:
\begin{eqnarray}
\hat{V}_{\rm ee} &=& (P_A+P_B) \hat{V}_{\rm ee} (P_A+P_B) \nonumber \\
&=& P_A \hat{V}_{\rm ee} P_A + P_B \hat{V}_{\rm ee} P_A 
+ P_A \hat{V}_{\rm ee} P_B + P_B \hat{V}_{\rm ee} P_B. \nonumber 
\end{eqnarray}
Here, $P_B \hat{V}_{\rm ee} P_B$ is not identical to $\hat{V}_{BB}$, where the expression 
of $\hat{V}_{BB}$ is given as a positive definite operator. 
The definition becomes explicit after the introduction of the orbital expansion. 
Similarly to $\hat{V}_{BB}$, we consider another partial interaction term, $\hat{V}_{AA}$, 
representing intra-correlation-band scattering processes. 
Two-body interaction processes by $P_A \hat{V}_{\rm ee} P_A$ consist of 
off-diagonal scattering processes by $\hat{V}_{AA}$, except for the
diagonal terms representing Hartree and exchange contributions of the core state 
$\hat{V}_{\rm core-core}$, and potential scattering terms $\hat{V}_{\rm core-corr}$ 
for electrons in correlated bands by the core electrons. 
We demonstrate that the Hubbard interaction is generated from $\hat{V}_{AA}$ 
as a restricted expansion of $\hat{V}_{\rm ee}$ by 
operators and wave functions in the correlated band 
that is screened by higher-order processes. 

By the definition of $\hat{V}_{BB}$, 
the scattering processes among high-energy quasi-particles (and/or quasi-holes) 
are deleted (or omitted) in the effective interaction of $\hat{V}_{\rm ee} - \hat{V}_{BB}$. 
This scheme involving reduced interactions allows 
us to define a closed set of relations (equations) 
among many-body Green's functions appearing in a later stage of our method. 
It should be noted that in the scattering process appearing in $P_B (\hat{V}_{\rm ee} - \hat{V}_{BB})P_B$, 
there must be at least one operator with an index in the band for strong correlation. 
After defining the orbital set $S_A$, 
the operator is specified as $c_{l,{\bf k},\sigma}^\dagger$ or $c_{l,{\bf k},\sigma}$ 
with an index $\{l,{\bf k}\} \in S_A$. 

The double-counting term $E_{\rm dc}^{(1)}[\Psi]$ is selected depending on 
$\hat{V}_{\rm ee}-\hat{V}_{BB}$. Namely, $E_{\rm dc}^{(1)}[\Psi]$ may be 
a mean-field contribution of $\langle \Psi| \hat{V}_{\rm ee}-\hat{V}_{BB} |\Psi\rangle$. 
It should be noted that the Hartree and exchange contributions of the core state 
do not appear in this term because 
$\langle \Psi_{\rm 0,core}| \hat{V}_{\rm ee}-\hat{V}_{BB} |\Psi_{\rm 0,core}\rangle =0$. 
When the mean-field contribution from interband scattering by 
$\hat{V}_{\rm ee}-\hat{V}_{BB}$ is treated as potential contribution adsorbed in the 
definition of $H_1$, if $\langle \Psi_{\rm 0,corr}| \hat{V}_{\rm ee}-\hat{V}_{BB} |\Psi_{\rm 0,corr}\rangle$
is approximated by a mean-field approximation of 
intra-correlated band scattering contribution $\langle \Psi| \hat{V}_{AA} |\Psi\rangle$, 
which is written as $E_{\rm intra-corr-MF}[n_\Psi]$, we can introduce 
a self-interaction-corrected Hartree approximation for this double-counting term. 
In any case of proper definition of the double-counting term, 
we consider that $E_{\rm dc}^{(1)}[\Psi]$ is given by a single-particle operator. 
Therefore, the functional derivative of $E_{\rm dc}^{(1)}[\Psi]$ with respect to 
$\langle\Psi|$ is a one-body operator defined as follows:
\begin{equation}
\hat{V}_{\rm dc}^{(1)}[\Psi] | \Psi \rangle 
\equiv \frac{\delta E_{\rm dc}^{(1)} [\Psi]}{\delta \langle\Psi|}. 
\end{equation} 

The determination equation of $|\Psi\rangle$ is given by the stationary condition, 
which is represented by a functional derivative of a Legendre-transformed functional 
with respect to $\langle \Psi|$, as follows:
\begin{eqnarray}
\lefteqn{\frac{\delta}{\delta \langle\Psi|}\left\{G_{\rm MR-DFT}[\Psi]
-E\left( \langle\Psi|\Psi\rangle -1 \right) \right\}} 
\nonumber \\
&=& \left\{ \hat{T} + V_{\rm ee} - V_{BB} - \hat{V}_{\rm dc}^{(1)} \right\} | \Psi \rangle 
\nonumber \\
&+&\int d^3r \left. \frac{\delta \left( E_{\rm Hartree}[n]+E_{\rm xc}[n] \right) }{\delta n({\bf r})}
\right|_{n=n_\Psi} \hat{n}({\bf r}) |\Psi\rangle \nonumber \\
&+& \int d^3r v_{\rm ext}({\bf r}) \hat{n}({\bf r}) |\Psi \rangle
-E |\Psi \rangle = 0. 
\label{MR-DFT_equation}
\end{eqnarray}
Another variational relationship by the derivative with respect to the 
multiplier $E$ gives the normalization of $|\Psi\rangle$ as $\langle \Psi | \Psi \rangle =1$. 
Therefore, the effective Hamiltonian is given by
\begin{eqnarray}
\hat{\cal H}_{\rm eff} &=& \hat{H}_1^{(0)} + \hat{H}_2^{(0)}, \\ 
\hat{H}_1^{(0)} &=& 
\hat{T} + \hat{V}_{\rm Hartree}[n_\Psi] + \hat{V}_{\rm xc}[n_\Psi] + \hat{V}_{\rm ext}, \\
\hat{H}_2^{(0)} &=& V_{\rm ee} - V_{BB}  - \hat{V}_{\rm dc}^{(1)}[\Psi]. 
\end{eqnarray}
Here, the one-body potential operators in $\hat{H}_1^{(0)}$ are as follows:
\begin{eqnarray} 
\hat{V}_{\rm Hartree}[n_\Psi] 
&=& \iint d^3r d^3r' \frac{e^2n_\Psi ({\bf r}')}{\left|{\bf r}-{\bf r}' \right|}\hat{n} ({\bf r}), \\
\hat{V}_{\rm xc}[n_\Psi] 
&=& \int d^3r 
\left[ \frac{\partial f}{\partial n} 
- \nabla \frac{\partial f}{\partial \nabla n} \right]_{n=n_\Psi} \hat{n} ({\bf r}), \\
\hat{V}_{\rm ext}
&=& \int d^3r v_{\rm ext}({\bf r}) \hat{n}({\bf r}). 
\end{eqnarray} 
The one-body Hamiltonian $\hat{H}_1^{(0)}$ is the Kohn-Sham Hamiltonian.  

An explicit representation of each operator 
is obtained by diagonalizing a single-particle part of the effective Hamiltonian.
To derive the effective one-particle potential, we can use the Kohn-Sham 
effective potential to determine the Kohn-Sham orbital. 
\begin{equation}
\hat{V}_{\rm eff}^{(0)} = \hat{V}_{\rm Hartree}[n_\Psi] + \hat{V}_{\rm xc}[n_\Psi] + \hat{V}_{\rm ext} .
\end{equation}
In the derivation of the single-particle orbital, we often include a one-particle operator 
of $\hat{V}_{\rm dc}^{(1)}[\Psi]$ in the effective single-particle operator as follows:
\begin{equation}
\hat{H}_1^{(1)} = 
\hat{T} + \hat{V}_{\rm Hartree} + \hat{V}_{\rm xc} + \hat{V}_{\rm ext} 
- \hat{V}_{\rm dc}^{(1)} .
\end{equation} 
In the LDA+U approach, a mean-field contribution from the two-body contribution 
of $\hat{V}_{\rm ee} - \hat{V}_{BB}$ may also be added to $\hat{H}_1$. 
Solutions of the single-particle orbital and their spectra differ 
depending on the definition of the single-particle effective Hamiltonian, $H_1$. 
\begin{equation}
H_1 \psi_{l, {\bf k}} ({\bf r})= \varepsilon_{l, {\bf k}} \psi_{l, {\bf k}}({\bf r}). 
\label{single_particle_eq}
\end{equation}
In addition, a residual part of the two-body interaction Hamiltonian $H_2 = H_{\rm eff} - H_1$ is redefined. 

The set of eigenfunctions $\{\psi_{l,{\bf k}}({\bf r})\}$ of $H_1$ expands $L^2(\Omega)$. 
When the set $\{\psi_{l,{\bf k}}({\bf r})\}$ is not complete, we can perform 
completion based on the completeness of the expanding numerical basis (e.g.,
plane-wave basis for a compact space). 
Here, for any integer $i$ within $i=1,\cdots, N$, 
a coordinate $(x_i,y_i,z_i,\sigma_i)$ is a point in $\Omega$, 
where $x_i$, $y_i$, and $z_i$ are in ${\mathbb R}$, and the two-dimensional spin coordinate 
is given by $\sigma=\uparrow$ or $\downarrow$. Therefore, we have a complete set 
to expand the Fock space of the electron system. 
Because we have an orthonormal basis set to define the field operators and 
electron creation (annihilation) operators, 
the effective many-body equation, 
\begin{equation} 
\hat{\cal H}_{\rm eff} |\Psi\rangle = E |\Psi\rangle, \label{many_body_eq}  
\end{equation}
is immediately expressed in a second-quantization form. 

When the diagonal basis of $H_1$ is used for the definition of creation (annihilation) 
operators $c^\dagger_{l,{\bf k},\sigma}$ ($c_{l,{\bf k},\sigma}$),  
\begin{equation}
\hat{H}_1 = \sum_{l,{\bf k}} \sum_{\sigma=\uparrow,\downarrow} \varepsilon_{l,{\bf k}} 
 c^\dagger_{l,{\bf k},\sigma} c_{l,{\bf k},\sigma}. 
\end{equation}
The Coulomb interaction is expanded as 
\begin{eqnarray}
\hat{V}_{\rm ee} &=& \frac{e^2}{2}
\sum_{l_i,{\bf k}_i,i=1,\cdots,4} \sum_{\sigma,\sigma'} \nonumber \\
&& \iint d^3r d^3r' \frac{
\psi^*_{l_1, {\bf k}_1}({\bf r}) \psi^*_{l_2, {\bf k}_2}({\bf r}')  
\psi_{l_3, {\bf k}_3}({\bf r}')\psi_{l_4, {\bf k}_4}({\bf r}) }{|{\bf r}-{\bf r}'|} \nonumber \\
&& c^\dagger_{l_1,{\bf k}_1,\sigma} c^\dagger_{l_2,{\bf k}_2,\sigma'} 
c_{l_3,{\bf k}_3,\sigma'} c_{l_4,{\bf k}_4,\sigma} . 
\label{Coulomb_expansion}
\end{eqnarray}
We introduce a set $S_A$ of orbital quantum numbers 
$\{l,{\bf k}\}$, where $\psi_{l,{\bf k}} ({\bf r}) $ is an orbital expanding the wave function of 
the correlated state $|\Psi_{\rm corr}\rangle$. 
The restricted scattering $\hat{V}_{AA}$ is defined as 
\begin{eqnarray}
\hat{V}_{AA} &=& \frac{e^2}{2}
\sum_{l_i,{\bf k}_i,i=1,\cdots,4,\{l_i,{\bf k}_i\}\in S_A} \sum_{\sigma,\sigma'} \nonumber \\
&& \iint d^3r d^3r' \frac{
\psi^*_{l_1, {\bf k}_1}({\bf r}) \psi^*_{l_2, {\bf k}_2}({\bf r}')  
\psi_{l_3, {\bf k}_3}({\bf r}')\psi_{l_4, {\bf k}_4}({\bf r}) }{|{\bf r}-{\bf r}'|} \nonumber \\
&& c^\dagger_{l_1,{\bf k}_1,\sigma} c^\dagger_{l_2,{\bf k}_2,\sigma'} 
c_{l_3,{\bf k}_3,\sigma'} c_{l_4,{\bf k}_4,\sigma} . 
\end{eqnarray}
Here, 
\begin{eqnarray}
\lefteqn{P_A \left( \hat{V}_{\rm ee} - \hat{V}_{BB} \right) P_A = P_A \hat{V}_{\rm ee} P_A} 
\nonumber \\ 
&=& \hat{V}_{AA} + \hat{V}_{\rm core-corr} + \hat{V}_{\rm core-core}. 
\end{eqnarray}

The complementary set $S_A^c$ of 
$S_A$ contains orbitals for core orbitals $S_A^{c,1}$ and empty orbitals $S_A^{c,2}$. 
We use $\hat{V}_{BB}$ as the next interaction term. 
\begin{eqnarray}
\hat{V}_{BB} &=& \frac{e^2}{2}
\sum_{l_i,{\bf k}_i,i=1,\cdots,4,\{l_i,{\bf k}_i\}\in S_A^c} \sum_{\sigma,\sigma'} \nonumber \\
&& \iint d^3r d^3r' \frac{
\psi^*_{l_1, {\bf k}_1}({\bf r}) \psi^*_{l_2, {\bf k}_2}({\bf r}')  
\psi_{l_3, {\bf k}_3}({\bf r}')\psi_{l_4, {\bf k}_4}({\bf r}) }{|{\bf r}-{\bf r}'|} \nonumber \\
&& c^\dagger_{l_1,{\bf k}_1,\sigma} c^\dagger_{l_2,{\bf k}_2,\sigma'} 
c_{l_3,{\bf k}_3,\sigma'} c_{l_4,{\bf k}_4,\sigma}  
\nonumber \\
&=& \iint d^3r d^3r' \sum_{\sigma,\sigma'}
\hat{A}_{\sigma,\sigma'}^\dagger({\bf r},{\bf r}') 
\frac{e^2}{2|{\bf r}-{\bf r}'|}
\hat{A}_{\sigma,\sigma'}({\bf r},{\bf r}') . \nonumber \\
\label{V_BB}
\end{eqnarray}
Here, the two-particle operator $\hat{A}_{\sigma,\sigma'}({\bf r},{\bf r}')$ is as follows:
\begin{eqnarray}
\lefteqn{\hat{A}_{\sigma,\sigma'}({\bf r},{\bf r}')} \nonumber \\
& = &
\sum_{l_1,{\bf k}_1,l_2,{\bf k}_2, \{l_i,{\bf k}_i\}\in S^c_A} 
\psi_{l_1, {\bf k}_1}({\bf r}) \psi_{l_2, {\bf k}_2}({\bf r}') 
c_{l_1,{\bf k}_1,\sigma} c_{l_2,{\bf k}_2,\sigma'}. \nonumber 
\end{eqnarray}
Because the Coulomb kernel is positive, and because $\hat{V}_{BB}$ 
consists of a linear combination of a positive form $g \hat{A}^\dagger \hat{A}$ ($g>0$), 
$\hat{V}_{BB}$ is positive definite. This characteristic allows us to apply 
the Kusakabe-Maruyama theorem for the converging model series of MR-DFT. 

Having the expanding basis by (\ref{single_particle_eq}), we can create an expansion of 
the minimizing state $|\Psi\rangle$ of $G_{\rm MR-DFT}[\Psi]$, 
which is the ground state determined by (\ref{many_body_eq}) as follows:
\begin{equation}
|\Psi\rangle = P_A |\Psi\rangle + P_B |\Psi\rangle .
\end{equation}
The A state of $|\Psi\rangle$ is given when 
\begin{equation}
P_A |\Psi\rangle \neq 0.
\end{equation} 
Practically, using the spectrum of (\ref{single_particle_eq}), we can define 
a core state $| \Psi_{\rm core}\rangle$ to have the A state in 
\begin{equation}
P_A |\Psi \rangle = |\Psi_{\rm corr} \rangle \otimes | \Psi_{\rm core}\rangle 
\end{equation}
and an expression of the core state as 
\begin{equation}
 | \Psi_{\rm core}\rangle  = \prod_{1\le l \le m, \; {\bf k}, \; {\rm s.t.} \; \varepsilon_{l,{\bf k}}<E_F} 
c^\dagger_{l,{\bf k},\uparrow} c^\dagger_{l,{\bf k},\downarrow} | 0 \rangle . 
\end{equation} 
It should be noted that a normalization constant appears in $|\Psi_{\rm corr}\rangle$. 
If necessary, we can renormalize the state vector so that the A state is treated as normalized. 

For the A state $|\Psi_A\rangle = P_A |\Psi_{\rm MF}\rangle$ 
of a mean-field description of LDA+U or random phase approximation (RPA), 
we can consider an exact exchange functional 
using the exchange density $n_{x,\Psi_A,\sigma}({\bf r},{\bf r}')$:
\begin{eqnarray}
\lefteqn{E_{\rm EXX} [n_{x,\Phi=\Psi_A,\sigma}]}\nonumber \\ 
&=& - \frac{1}{2} \iint d^3rd^3r' \sum_\sigma \frac{e^2}{\left|{\bf r}-{\bf r}' \right|}
n_{x,\Phi,\sigma}({\bf r},{\bf r}')n_{x,\Phi,\sigma}({\bf r}',{\bf r}). \nonumber \\
\end{eqnarray}
The form of the Fock-Dirac density matrix 
$n_{x,\Phi,\sigma}({\bf r},{\bf r}')$ is given by the orbital 
$\left\{\psi_{l,{\bf k}}({\bf r})\right\}$ as eigenfunctions with orbital energies 
$\varepsilon_{l,{\bf k}}$ of a single-particle part of the effective Hamiltonian. 
Because the Fermi level $E_F$ is determined as 
the highest occupied level of the A state for LDA+U or RPA, 
and because a paramagnetic ground state is considered, the reduced density matrix is written as follows: 
\begin{equation}
n_{x,\Phi,\sigma}({\bf r},{\bf r}')  = \sum_{\varepsilon_{l,{\bf k}} \leq E_F } 
\psi^*_{l,{\bf k}}({\bf r}) \psi_{l,{\bf k}}({\bf r}').  \nonumber \\
\end{equation}
However, the functional $E_{\rm EXX}$ is ill defined for $|\Psi_{\rm MF}\rangle$ 
when the B state $|\Psi_B\rangle = (\hat{1}-P_A)|\Psi_{\rm MF}\rangle$ is relevant. 

By letting $P_A \rightarrow \hat{I}$, $P_B \rightarrow \hat{0}$, $\hat{V}_{BB} \rightarrow \hat{0}$, 
$E_{dc}^{(1)}[\Psi] \rightarrow E_{\rm Hartree} [n_\Psi]$, and 
$E_{\rm xc}[n_\Psi] \rightarrow 0$, 
(or $E_{dc}^{(1)}[\Psi] \rightarrow E_{\rm Hartree} [n_\Psi] + E_{\rm xc}[n_\Psi]$), 
$G_{\rm MR-DFT}[\Psi]$ can represent 
the Coulomb energy functional. 
Other choices for the double-counting term and exchange-correlation functional are 
discussed after introducing LDA+U or RPA. 
When $E_{\rm fluc} [\Psi]=0$, $G_{\rm MR-DFT}[\Psi]$ becomes the Kohn-Sham energy functional 
in the exact functional form of the wave function \cite{Hadjisavvas}.
The minimizing state $|\Psi\rangle$ becomes the Kohn-Sham state. 

Here, we derive an effective Hamiltonian to determine the A state. 
First, we rewrite (\ref{many_body_eq}) in separated forms 
assuming that the A state is determined by the Kohn-Sham orbitals given by $H_1^{(0)}$ as follows:
\begin{eqnarray}
\lefteqn{
\left\{ \hat{T} + \hat{V}_{\rm eff}^{(0)}\right\} P_A |\Psi\rangle  + 
P_A\left(\hat{V}_{\rm ee} -\hat{V}_{\rm dc}^{(1)} \right) P_A |\Psi\rangle }
\nonumber \\
&+ P_A \left(\hat{V}_{\rm ee} -\hat{V}_{\rm dc}^{(1)}\right) P_B |\Psi\rangle = E P_A |\Psi\rangle , 
\label{A_state_equation}\\
\lefteqn{
\left\{ \hat{T} + \hat{V}_{\rm eff}^{(0)} \right\} P_B |\Psi\rangle} \nonumber \\
&+P_B\left(\hat{V}_{\rm ee} - \hat{V}_{BB}-\hat{V}_{\rm dc}^{(1)} \right) P_B |\Psi\rangle 
\nonumber \\
&+ P_B \left(\hat{V}_{\rm ee} -\hat{V}_{\rm dc}^{(1)}\right) P_A |\Psi\rangle = E P_B |\Psi\rangle .
\end{eqnarray}
Here, $\hat{V}_{\rm dc}^{(1)}$ causes off-diagonal potential scattering for $P_A |\Psi \rangle$, 
where $H_1^{(0)}$ is diagonal with respect to $P_A |\Psi \rangle$. In this expression, 
$P_B|\Psi\rangle$ is determined as 
\begin{eqnarray}
\lefteqn{P_B|\Psi\rangle } \nonumber \\
&=& \left\{ E - \hat{T} - \hat{V}_{\rm eff}^{(0)} -
P_B\left(\hat{V}_{\rm ee} - \hat{V}_{BB}-\hat{V}_{\rm dc} \right)P_B\right\}^{-1}  \nonumber \\
&\times&
P_B \left(\hat{V}_{\rm ee} -\hat{V}_{\rm dc}^{(1)}\right) P_A |\Psi\rangle. 
\label{B_state_expression1}
\end{eqnarray}
Inserting (\ref{B_state_expression1}) into (\ref{A_state_equation}), 
we obtain the determination equation of $P_A|\Psi\rangle$. 

When we adopt an effective potential including $\hat{V}_{\rm dc}^{(1)}$, 
\begin{equation}
\hat{V}_{\rm eff}^{(1)} = \hat{V}_{\rm Hartree}[n_\Psi] + \hat{V}_{\rm xc}[n_\Psi] + \hat{V}_{\rm ext} 
- \hat{V}_{\rm dc}^{(1)}, 
\end{equation}
we obtain another expression for the separated form. 
\begin{eqnarray}
\lefteqn{
\left\{ \hat{T} + \hat{V}_{\rm eff}^{(1)} + 
P_A\hat{V}_{\rm ee}\right\} P_A|\Psi\rangle }
\nonumber \\
&+ P_A \hat{V}_{\rm ee}P_B |\Psi_B\rangle = E P_A|\Psi\rangle , \\
\lefteqn{
\left\{ \hat{H}_1 +
P_B\left(\hat{V}_{AA} + \hat{V}_{B\leftrightarrow A} \right)\right\} P_B|\Psi\rangle}
\nonumber \\
&+ P_B \hat{V}_{\rm ee} P_A |\Psi\rangle = E P_B |\Psi\rangle .
\label{B_state_equation2}
\end{eqnarray}
In the above equations, we use another interaction operator, 
$\hat{V}_{B\leftrightarrow A}$, 
which is defined by the following relations. 
\begin{eqnarray}
\hat{V}_{\rm ee} &=& \hat{V}_{AA} + \hat{V}_{BB} + \hat{V}_{B\leftrightarrow A}.
\end{eqnarray}
The expression of (\ref{B_state_equation2}) leads to another form of $P_B |\Psi\rangle$. 
\begin{eqnarray}
P_B |\Psi\rangle 
&=& P_B \left\{ E - \hat{H}_1 - \hat{V}_{AA} - \hat{V}_{B\leftrightarrow A} \right\}^{-1}  \nonumber \\
&\times& P_B \hat{V}_{\rm ee} P_A |\Psi\rangle. \label{B_state_expression}
\end{eqnarray}

Using (\ref{B_state_expression}), 
we can derive an effective Hamiltonian for $|\Psi_A\rangle = P_A |\Psi\rangle$ as follows:
\begin{eqnarray}
\lefteqn{\hat{\cal H}_{\rm red} (E) } \nonumber \\
&=& \hat{T} + \hat{V}_{\rm Hartree}[n_\Psi] + \hat{V}_{\rm xc}[n_\Psi] + \hat{V}_{\rm ext} 
+ P_A \hat{V}_{\rm ee} P_A \nonumber \\
&-&P_A \hat{V}_{\rm ee} P_B 
\left\{\hat{H}_1 + \hat{V}_{AA} + \hat{V}_{B\leftrightarrow A} - E \right\}^{-1} 
\nonumber \\
&\times& P_B \hat{V}_{\rm ee} P_A 
- \hat{V}_{\rm dc}^{(1)} . \label{A_state_Hamiltonian}
\end{eqnarray}
\begin{equation}
\hat{\cal H}_{\rm red} (E) |\Psi_A\rangle = E |\Psi_A\rangle.
\end{equation}

In this model Hamiltonian of (\ref{A_state_Hamiltonian}), there is a resolvent operator, 
which is a generalized Green function. 
\begin{eqnarray}
\lefteqn{G^{(2)}(\omega)} \nonumber \\
&=& \left\{\omega - \left(\hat{H}_1 + \hat{V}_{AA} + \hat{V}_{B\leftrightarrow A} \right) \right\}^{-1}
\end{eqnarray}

In the resolvent operator $G^{(2)}(\omega)$, 
there is a source term $\hat{V}_{B\leftrightarrow A}$ creating a high-energy excited state 
via electron-hole pair creation. 
When $G^{(2)}(\omega)$ is expanded in a series of 
$G^{(2,0)}(\omega) =\left\{ \omega - \hat{H}_1 - \hat{V}_{AA} \right\}^{-1}$, 
there appears a high-energy state $|\Psi_{l,N_{\rm corr}=0}\rangle$ 
with no electrons in the correlated bands; index $l$ labels these high-energy states. 
Then, $\hat{V}_{AA}$ causes no scattering event, and the elements 
of $G^{(2,0)}(\omega)$ with respect to $|\Psi_{l,N_{\rm corr}=0}\rangle$ are decomposed 
into two single-particle Green's functions. 
Therefore, our effective model is solvable in the sense that 
finiteness in the determination equations of its Green's function is certified. 
This property arises from the form of $\hat{V}_{\rm ee} - \hat{V}_{BB}$. 

To consider $\lambda$ integration in the following sections, 
we introduce $\lambda$-modified models. 
In the model space defined by $\hat{\cal H}_{\rm eff}$, 
$\hat{\cal H}_{\rm red} (E)$, and their variants, 
we have $\lambda$ integration paths. 
As an example, we consider the next $\lambda$-modified Hamiltonian, 
$\hat{\cal H}_\lambda^{(0)}$:
\begin{equation}
\hat{\cal H}_\lambda^{(0)} (E)
= \lambda \hat{\cal H}_{\rm Coulomb} + (1-\lambda) \hat{\cal H}_{\rm eff}.
\label{lambda_modified_0}
\end{equation}
The domain of $\hat{\cal H}_{\rm eff}$ is the same as that of $\hat{\cal H}_{\rm Coulomb}$. 
When $P_A\rightarrow \hat{I}$, $\hat{V}_{BB}\rightarrow \hat{0}$, 
if $E_{dc}^{(1)}[\Psi]\rightarrow E_{\rm Hartree}[n_\Psi]+E_{\rm xc}[n_\Psi]$, 
the stable ground state $|\Psi\rangle$ of $\hat{\cal H}_{\rm eff}$ approaches 
$|\Psi_0\rangle$ of $\hat{\cal H}_{\rm Coulomb}$. 
If $|\Psi_0\rangle$ has a corresponding A state as $P_A|\Psi_0\rangle\neq 0$, 
we have a well-defined $\lambda$ integration path connecting 
$\hat{\cal H}_\lambda^{(1)} (E)$ and $\hat{\cal H}_{\rm Coulomb}$ with a seamless 
state vector being the lowest-energy eigenstate. 
Then, the next $\lambda$-modified effective Hamiltonian exists. 
\begin{equation}
\hat{\cal H}_\lambda^{(1)} (E)
= \lambda \hat{\cal H}_{\rm Coulomb} + (1-\lambda) \hat{\cal H}_{\rm red}(E).
\label{lambda_modified_1}
\end{equation}
When $\lambda=1$, this model Hamiltonian is identical to $\hat{\cal H}_{\rm Coulomb}$. 

\subsection{Reduced Hubbard-type Hamiltonian} 

Here, we define a Hubbard model as a variant of $\hat{\cal H}_{\rm red}(E)$. 
Let us consider a form similar to the reduced Hamiltonian as follows:
\begin{eqnarray}
\lefteqn{\hat{\cal H}_{\rm Hubbard} } \nonumber \\
&=& \hat{T} + \hat{V}_{\rm Hartree}[n_{\rm Hubbard}]
+ \hat{V}_{\rm xc}[n_{\rm Hubbard}] \nonumber \\
&& 
+ \hat{V}_{\rm Hubbard} -\hat{V}_{\rm dc}^{(2)} + \hat{V}_{\rm ext}.
\label{DFT_Hubbard}
\end{eqnarray}
We have a model A state, which is the ground state of the Hubbard Hamiltonian. 
\begin{equation}
\hat{\cal H}_{\rm Hubbard} |\Psi_{A} \rangle = E |\Psi_{A}\rangle .
\end{equation}
As discussed, the A state appears with a corresponding B state in the following form: 
\begin{eqnarray}
|\Psi_B\rangle 
&=& P_B G^{(2)}(\omega =E) 
P_B \hat{V}_{\rm int} P_A |\Psi_A\rangle. 
\end{eqnarray}
In this definition, $\hat{V}_{\rm ee}$ is replaced by a model two-body interaction, $\hat{V}_{\rm int}$. 
This is because the scattering process must be restricted within an effective on-site interaction 
in the definition of the Hubbard model. 
In addition, the energy $E$ appearing in the resolvent is approximated by the eigenvalue of 
the resulting model. This treatment is justified for some cases, which are discussed later. 
With the B state, the full state vector of the stable state of the Hubbard model is as follows:
\begin{eqnarray}
|\Psi_{\rm Hubbard}\rangle &=& |\Psi_A\rangle + |\Psi_B\rangle, \\
|\Psi_A\rangle &=& P_A |\Psi_{\rm Hubbard}\rangle, \nonumber \\
|\Psi_B\rangle &=& P_B |\Psi_{\rm Hubbard}\rangle, \nonumber \\
n_{\rm Hubbard}({\bf r}) &=& \langle\Psi_{\rm Hubbard}|\hat{n}({\bf r}) | \Psi_{\rm Hubbard}\rangle. 
\end{eqnarray}
Due to the virtual excitations represented by $|\Psi_B\rangle$, the effective model 
is described by the Hubbard short-range interaction with the screened interaction 
parameter. An explicit form of the screened Coulomb interaction is provided in the following section. 

As a typical example, we consider a single-band model, that is, a correlated band 
with a fixed band index, $n$. The Wannier transformation is introduced as a Fourier transformation 
of the Bloch state. The electron creation (and annihilation) operators $c_{n,i,\sigma}^\dagger$ 
($c_{n,i,\sigma}$) in the site representation are given by the Wannier orbitals, and 
the number operator is given as $n_{n,i,\sigma} = c_{n,i,\sigma}^\dagger c_{n,i,\sigma}$. 
We use the average $\bar{n}_{n,i,\sigma} = \langle \Psi_{\rm Hubbard} | n_{n,i,\sigma} |\Psi_{\rm Hubbard} \rangle$. 
The on-site interaction term can be introduced as the effective model interaction.
The double-counting term is introduced considering the Hartree contribution 
for the local-orbital correlation with the self-interaction correction for the 
Hartree-type mean-field term. 
\begin{eqnarray}
\lefteqn{\hat{V}_{\rm Hubbard} -\hat{V}_{\rm dc}^{(2)} } \nonumber \\ 
&=& \frac{1}{2} \sum_{i} U_n \left\{
:\left( \hat{n}_{n,i,\uparrow}+\hat{n}_{n,i,\downarrow} - \bar{n}_{n,i,\uparrow}-\bar{n}_{n,i,\downarrow}\right)^2: \right. \nonumber \\
&& \left.
+ \left( \bar{n}_{n,i,\uparrow} + \bar{n}_{n,i,\downarrow}\right)\right\}
\end{eqnarray}
This choice of $\hat{V}_{\rm dc}^{(2)}$ is the same as the well-known choice of 
the double-counting term \cite{Anisimov,Anisimov2,Solovyev}.
To omit the interaction term in the on-site Hubbard interaction, the scattering processes 
caused by $\hat{V}_{AA}$ and $\hat{V}_{\rm int}$ 
are limited so that only the site-diagonal terms appear. 

We can consider several $\lambda$-integration paths connecting relevant models. 
As an example, we introduce a path from the Hubbard-type model to the Coulomb model. 
Because the domain of (\ref{DFT_Hubbard}) is the same as the domain 
of $\hat{\cal H}_{\rm Coulomb}$, we can obtain the next $\lambda$-modified effective Hamiltonian. 
\begin{equation}
\hat{\cal H}_\lambda^{(2)} 
= \lambda \hat{\cal H}_{\rm Coulomb} + (1-\lambda) \hat{\cal H}_{\rm Hubbard}.
\label{lambda_modified_2}
\end{equation}
This integration path may have a singular point, which is in part because 
$|\Psi_{\rm Hubbard}\rangle$ represents limited types of strong correlation effects. 
Therefore, we should consider a path bypassing the point. 
\begin{equation}
\hat{\cal H}_\lambda^{(3)} 
= \lambda \hat{\cal H}_{\rm red}(E) + (1-\lambda) \hat{\cal H}_{\rm Hubbard}.
\label{lambda_modified_3}
\end{equation}
Along the path, off-site interactions are recovered. Then, we can analyze 
the possible phase transition points within the effective interaction model with 
the A state representation. 

\subsection{RPA evaluation of $U_n$}

The many-particle Green's function $G^{(2)}(\omega =E)$ in $\hat{\cal H}_{\rm red}(E)$ 
is a resolvent operator, and
$G^{(2)}(\omega =E)$ generally behaves as a many-particle Green's function. 
Because $P_B\hat{V}_{\rm ee}P_A$ acting on $P_A|\Psi\rangle$ 
creates two electron-hole pairs in the normal sense, 
there appears a two-particle Green's function in the effective interaction of 
\begin{eqnarray}
\lefteqn{\hat{W}_{\rm red}(\omega) \equiv  P_A \hat{W}(\omega) P_A} \nonumber \\
&=& P_A\hat{V}_{\rm ee}P_A + P_A\hat{V}_{\rm ee}P_BG^{(2)}(\omega) P_B\hat{V}_{\rm ee}P_A.
\end{eqnarray}
The energy variable $\omega=E$ is a parameter that should be equated as the eigenenergy 
when searching for the eigenstate $|\Psi\rangle$. 
By treating $\omega$ as a variable, we can derive several approximation methods. 

With an approximation on a final state $|\Psi_{B}\rangle = P_B\hat{V}_{\rm ee}P_A|\Psi\rangle$, 
we can introduce a static approximation on the screened interaction. 
In general, $|\Psi_{B}\rangle$ has several quasi-electrons (and/or quasi-holes) 
in the $S_A^{c,1}$ (and/or $S_A^{c,2}$) space.
When we decompose it into $|\Psi_{B}\rangle=|\Psi_{B,corr}\rangle \otimes |\Psi_{B,S_A^c}\rangle$, 
the electronic state $|\Psi_{B,corr}\rangle$ in the correlated band has a recoil effect. 
Typically, when the correlated state suffers from a short-range correlation, 
the creation of the resulting double occupancy of electrons (or holes) should result in an energy enhancement. 
We represent this effect by energy variable $U_{\rm eff}$. Assuming $U_{\rm eff}$ to be constant, 
we have a type of static approximation. 
Setting the origin of the frequency variable $\omega$ to zero as the resulting 
eigenenergy, which is accepted by neglecting the energy of $|\Psi_{B,corr}\rangle$ in 
the energy denominator, we obtain an effective static screened interaction operator. 
Considering the relevant contribution from the two-particle Green's function, 
which is described by a response operator $\hat{\chi}(\omega)$, 
and assuming that the vertex correction is negligible, we have the following RPA expression. 
\begin{eqnarray}
\lefteqn{\hat{W}_{\rm red} }\nonumber \\
&=& 
P_A \hat{W}(\omega=0^+) P_A \nonumber \\
&\simeq& 
P_A\left(\hat{V}_{\rm ee} + \hat{V}_{\rm ee} \hat{\chi}(\omega=0^+) 
\frac{1}{1-\hat{V}_{\rm ee} \hat{\chi}(\omega=0^+)} 
\hat{V}_{\rm ee}\right)P_A \nonumber \\
&=&
P_A\left(1-\hat{V}_{\rm ee} \hat{\chi}(\omega=0^+)\right)^{-1}\hat{V}_{\rm ee}P_A,
\end{eqnarray}
where 
\begin{equation}
\chi ({\bf r},{\bf r}';\omega)
=P_{(c,p)}({\bf r},{\bf r}';\omega) + P_{(p,v)}({\bf r},{\bf r}';\omega).
\end{equation}
The polarization functions are effectively expressed as
\begin{eqnarray}
\lefteqn{P_{(c,p)}({\bf r},{\bf r}';\omega)}\nonumber \\
&=&
2\sum_{i=(n_1,{\bf k}_1)\in S_A}\sum_{j=(n_2,{\bf k}_2)\in S_A^{c,2}}
\phi^*_i({\bf r})\phi_j({\bf r})\phi^*_j({\bf r}')\phi_i({\bf r}')
\nonumber \\
&\times&\left[\frac{1}{\omega-E_j-U_{\mathrm{eff}}+E_i+i\delta}-\frac{1}{\omega+E_j+U_{\mathrm{eff}}-E_i-i\delta}\right], \nonumber \\
\label{eq:mr-dft_1}
\end{eqnarray}
\begin{eqnarray}
\lefteqn{P_{(p,v)}({\bf r},{\bf r}';\omega)}\nonumber \\
&=&2\sum_{i=(n_1,{\bf k}_1)\in S_A^{c,1}}\sum_{j=(n_2,{\bf k}_2)\in S_A}
\phi^*_i({\bf r})\phi_j({\bf r})\phi^*_j({\bf r}')\phi_i({\bf r}') \nonumber \\
&\times&\left[\frac{1}{\omega-E_j-U_{\mathrm{eff}}+E_i+i\delta}-\frac{1}{\omega+E_j+U_{\mathrm{eff}}-E_i-i\delta}\right]. \nonumber \\
\label{eq:mr-dft_2}
\end{eqnarray}

The static assumption on $\omega \simeq 0$ is accepted in several cases. 
When there is an energy gap between states in $S_A^{c,1}$ and $S_A^{c,2}$, 
if $U_{\rm eff}$ is nearly the same as the constant for the relevant excitations, 
the above expressions, (\ref{eq:mr-dft_1}) and (\ref{eq:mr-dft_2}), for the polarization functions should be effective. 
If we further assume that $U_{\rm eff}$ is negligible, when we approximate 
$\chi(\omega)$ by an RPA estimation of the response function, we can obtain 
the constrained RPA (cRPA) calculation given by the following form \cite{cRPA}:
\begin{equation}
W_{\rm cRPA}
=\left(1-{V}_{\rm ee} {\chi}_{\rm RPA}(\omega=0^+)\right)^{-1}{V}_{\rm ee}.
\end{equation}

When the c-RPA-screened interaction is used to evaluate the on-site Hubbard interaction, 
we obtain the following formula:
\begin{eqnarray}
U_n^{(r)} &=& \left(n,i,n,i \right|  W_{\rm cRPA}(\omega=0^+)\left|n,i,n,i \right) \nonumber \\
&=&\iint d^3rd^3r' \phi_{n,i}^*({\bf r}) \phi_{n,i}^*({\bf r}')  \nonumber \\
&& \times W_{\rm cRPA}({\bf r},{\bf r}';\omega=0^+)
\phi_{n,i}({\bf r}')\phi_{n,i}({\bf r}). 
\end{eqnarray}
Using the cRPA form, we obtain the Hubbard interaction with its double-counting correction term 
as follows:
\begin{eqnarray}
\lefteqn{\hat{V}_{\rm Hubbard}^{(r)}[\Psi_{\rm Hubbard}] -\hat{V}_{\rm dc}^{(r,2)}[\Psi_{\rm Hubbard}]} \nonumber \\ 
&=& \sum_{i} U_n^{(r)} \left\{ \hat{n}_{n,i,\uparrow} \hat{n}_{n,i,\downarrow} 
-\bar{n}_{n,i}\left( \hat{n}_{n,i,\uparrow}+\hat{n}_{n,i,\downarrow} \right) \right\}
\nonumber \\
&& + \frac{1}{2} \sum_{i} U_n^{(r)} 
\bar{n}_{n,i}\left( \bar{n}_{n,i} + 1\right).
\end{eqnarray}
Following the standard definition of the Hubbard interaction, we define the following separation 
of the fluctuation terms. 
\begin{eqnarray}
\hat{V}_{\rm Hubbard}^{(r)}[\Psi] 
&=&\sum_{i} U_n^{(r)} \hat{n}_{n,i,\uparrow} \hat{n}_{n,i,\downarrow} , \\
\hat{V}_{\rm dc}^{(r,2)}[\Psi]
&=&\sum_{i} U_n^{(r)}\bar{n}_{n,i}\left( \hat{n}_{n,i,\uparrow}+\hat{n}_{n,i,\downarrow} \right) 
\nonumber \\
&&-\frac{1}{2} \sum_{i} U_n^{(r)} 
\bar{n}_{n,i}\left( \bar{n}_{n,i} + 1\right).
\end{eqnarray}

\subsection{Mean-field approximation as the LDA+U approximation}

In the proposed theory of LDA+U approximation, we consider that i) $U_{n}^{(r)}$ is given by 
a multi-configuration state whose A state takes the form of the mean-field state 
(a single Slater determinant), and ii) the renormalized interaction determines 
a unique mean-field ground state. 
The first condition is natural and is similar to cRPA. 
In the RPA expansion of the many-body perturbation theory, 
the LDA ground state is used to create the 0th order Green function. 
The second assumption is validated by deriving a numerical solution. 
Practically, the known mean-field LDA+U solver is adopted having $U_n^{(r)}$ by cRPA. 

In this section, we derive several formal expressions on the LDA+U approach. 
Initially, our mean-field ground state has both A-state and B-state parts, 
as the B state expresses the screening. 
\begin{equation}
|\Phi_{\rm MF}\rangle = |\Phi_A\rangle + |\Phi_B\rangle.
\end{equation}
At the same time, the mean-field treatment of the Hubbard model allows us to write 
the A state vector in the form of a single Slater. It is divided into a direct product form. 
\begin{equation}
|\Phi_A\rangle = |\phi_{\rm MF} \rangle \otimes | \Phi_{\rm core}\rangle 
\end{equation}
The core state is written in the following form. 
\begin{equation}
 | \Phi_{\rm core}\rangle  = \prod_{1\le l <n, \; {\bf k}, \; {\rm s.t.} \; \varepsilon_{l,{\bf k}}<E_F} 
c^\dagger_{l,{\bf k},\uparrow} c^\dagger_{l,{\bf k},\downarrow} | 0 \rangle . 
\end{equation} 
In addition, for the correlated $n$th band, the mean-field approach allows us to write the stationary state 
in the following Fermi sea. 
\begin{equation}
 | \phi_{\rm MF}\rangle  = \prod_{{\bf k}, \; {\rm s.t.} \; \varepsilon_{n,{\bf k}}<E_F} 
c^\dagger_{n,{\bf k},\uparrow} c^\dagger_{n,{\bf k},\downarrow} | 0 \rangle . 
\end{equation} 
The average value of the number operators is as follows: 
\begin{equation}
\bar{n}_{n,i,\sigma}
=\langle \Phi_{\rm MF} | \hat{n}_{n,i,\sigma}| \Phi_{\rm MF}\rangle.
\end{equation}
Here, we define the expectation value of the total electron number at the $i$th site. 
\begin{equation}
\bar{n}_{n,i}=\bar{n}_{n,i,\uparrow} +\bar{n}_{n,i,\downarrow}.
\end{equation}
Now, the expectation value of the Hubbard interaction is counted 
within the mean-field approximation using the electron occupation at the $i$th site. 
\begin{eqnarray}
\lefteqn{\bar{E}_{\rm Hubbard} 
= \langle \Phi_{\rm MF} | \hat{V}_{\rm Hubbard}^{(r)} | \Phi_{\rm MF}\rangle}\nonumber \\
&=& U_n^{(r)} \sum_{i} \bar{n}_{n,i,\uparrow} \bar{n}_{n,i,\downarrow} .
\end{eqnarray}
It is also given in the momentum representation. 
\begin{eqnarray}
\bar{E}_{\rm Hubbard} &=& \frac{1}{2}\sum_{{\bf k}_1,{\bf k}_2} \sum_\sigma 
U_n^{(p)} \langle \hat{n}_{n,{\bf k}_1,\sigma} \rangle \langle \hat{n}_{n,{\bf k}_2,-\sigma} \rangle .
\end{eqnarray}
In this expression, the interaction strength is given by 
\begin{equation}
U_n^{(p)} = 
\left[\frac{1}{N^4}\sum_{{\bf p}_1,{\bf p}_2,{\bf p}_3,{\bf p}_4} 
\left(n,{\bf p}_1,n,{\bf p}_2 \right| W_{\rm cRPA} \left|n,{\bf p}_3,n,{\bf p}_4 \right) \right].
\label{U_in_momentum}
\end{equation}
Therefore, the delta-type on-site interaction has an average value of interaction strength (i.e., average scattering amplitude). 

The mean-filed contribution of the double-counting term reads
\begin{equation}
\bar{E}_{\rm dc}^{(3)} = 
\langle \Phi_{\rm MF} | \hat{V}_{\rm dc}^{(r,2)} | \Phi_{\rm MF}\rangle .
\end{equation}
It is explicitly given as follows:
\begin{eqnarray}
\lefteqn{\bar{E}_{\rm dc}^{(3)} }\nonumber \\
&=& 
\frac{U_n^{(r)}}{2} \sum_{i} (\bar{n}_{n,i,\uparrow}+ \bar{n}_{n,i,\downarrow} )
(\bar{n}_{n,i,\uparrow} + \bar{n}_{n,i,\downarrow} -1).
\end{eqnarray}
We can obtain the expectation value of the Hartree energy 
and exchange-correlation energy by calculating 
the charge density 
\begin{equation}
n_{\rm MF}({\bf r}) = \langle\Phi_{\rm MF} |\hat{n}({\bf r})| \Phi_{\rm MF}\rangle. 
\end{equation} 
The total energy in the LDA+U approximation is given by the following formula:
\begin{eqnarray}
E_{\rm LDA+U} &=& E_{\rm kin} + E_{\rm Hartree}[n_{\rm MF}] 
+ E_{\rm xc}[n_{\rm MF}] \nonumber \\ 
&+& \bar{E}_{\rm Hubbard} - \bar{E}_{\rm dc}^{(3)} + \bar{E}_{\rm ext}. 
\end{eqnarray} 
Next, the effective potential part for the LDA+U model is considered. 
The mean-field operator of the Hubbard interaction is as follows:
\begin{equation}
\hat{\bar{V}}_{\rm Hubbard} 
= \sum_{i} U_n^{(r)} 
\left\{ 
\bar{n}_{n,i,\downarrow} \hat{n}_{n,i,\uparrow} 
+\bar{n}_{n,i,\uparrow} \hat{n}_{n,i,\downarrow} 
\right\}
\end{equation}
The double-counting term produces the following single-particle operator, which 
is also a potential scattering term. 
\begin{equation}
\hat{\bar{V}}_{\rm dc}^{(3)}  
= \frac{U_n^{(r)}}{2} \sum_{i}  \left(2\bar{n}_{n,i} - 1\right) 
\left( \hat{n}_{n,i,\uparrow}+\hat{n}_{n,i,\downarrow} \right) .
\end{equation}

In the formal theory above, the ground state $|\Phi_{MF}\rangle$ is a multi-reference state. 
Once the mean-field Hamiltonian is given using the effective operators, 
we can treat only the A state, $|\Phi_A\rangle$. 
In the calculation, 
the normalized state vector $|\tilde{\Phi}_{A}\rangle=C_{LDA+U}|\Phi_A\rangle$ is explicitly given. 
However, it is difficult to obtain the B state, $|\Phi_B\rangle$.  
Usually, a normalization condition of $\langle\tilde{\Phi}_{A}|\tilde{\Phi}_{A}\rangle =1$ is adopted, and 
the normalization constant $C_{LDA+U}$ is not explicitly determined. 
Therefore, the correspondence of 
the LDA+U ground state is as follows:
\begin{equation}
|\Phi_{\rm LDA+U}\rangle = |\tilde{\Phi}_{A}\rangle .
\end{equation}
Furthermore, an approximation is generally used for the charge density 
and occupation numbers as 
\begin{eqnarray}
n_{\rm LDA+U}({\bf r}) &=& \langle\Phi_{\rm LDA+U} |\hat{n}({\bf r})| \Phi_{\rm LDA+U}\rangle. \\
\tilde{n}_{n,i,\sigma} &=& \langle\Phi_{\rm LDA+U} | \hat{n}_{n,i,\sigma}| \Phi_{\rm LDA+U}\rangle.
\end{eqnarray}
The determination equation of $|\tilde{\Phi}_{A}\rangle$ is as follows:
\begin{eqnarray}
&&\left\{\hat{T} + \hat{V}_{\rm Hartree}[n_{\rm LDA+U}]
+\hat{V}_{\rm xc}[n_{\rm LDA+U}] \right. \nonumber \\ 
&&\left.
+ \hat{\tilde{V}}_{\rm Hubbard} - \hat{\tilde{V}}_{\rm dc}^{(3)}  
+\hat{V}_{\rm ext} \right\}|\tilde{\Phi}_{A}\rangle \nonumber \\
&&=E|\tilde{\Phi}_{A}\rangle.
\end{eqnarray}
Operators $\hat{\tilde{V}}_{\rm Hubbard}$ and $\hat{\tilde{V}}_{\rm dc}^{(3)}$ 
are defined using $\tilde{n}_{n,i,\sigma}$, and 
the LDA+U Hamiltonian is given by
\begin{eqnarray}
\hat{H}_1^{LDA+U}&=&
\hat{T} + \hat{V}_{\rm Hartree}[n_{\rm LDA+U}]
+\hat{V}_{\rm xc}[n_{\rm LDA+U}]  \nonumber \\
&+& \hat{\tilde{V}}_{\rm Hubbard} - \hat{\tilde{V}}_{\rm dc}^{(3)}  
+\hat{V}_{\rm ext}  . 
\end{eqnarray}
Because the problem is given as a single-particle picture, we have the following LDA+U equation:
\begin{equation}
H_1^{LDA+U} \phi_{l, {\bf k}} ({\bf r})= \varepsilon^{LDA+U}_{l, {\bf k}} \phi_{l, {\bf k}}({\bf r}). 
\end{equation}

Here, two points should be noted. 
First, the set of eigenfunctions of $H_1^{LDA+U}$ also expands $L^2(\Omega)$; 
that is, ${\rm dim} \{|\Psi_A\rangle \} = {\rm dim} \{|\Phi_A\rangle \}$. 
Therefore, we can use $\phi_{l, {\bf k}} ({\bf r})$ and $\varepsilon^{LDA+U}_{l, {\bf k}}$ 
to construct many-body perturbation theory similar to other Kohn-Sham orbitals. 

In the approximation, a single reference state of $|\Phi_A\rangle$ appears. 
This state and its orbital components are different from LDA/GGA Kohn-Sham orbitals. 
As discussed in the main text of this paper, even
the charge density $n_{\rm LDA+U}({\bf r})$ is different from the density by LDA. 
As a DFT, a relevant point causing the difference between approximations is 
the charge density. In comparison to LDA, for example, the determination equation of 
the orbital is modified by the inclusion of screened interaction effects in a partially occupied 
correlated band around the Fermi energy, $E_F$. 
In (\ref{U_in_momentum}), it can be seen that the short-range interaction part caused by 
$E_{\rm fluc}$ is described by a momentum-averaged screened Coulomb interaction. 
This part contains a specific scattering effect explicitly out of the other scattering 
contribution evaluated later by RPA. 

When $P_A$ and $E_{\rm fluc}$ 
are properly selected, the MR-DFT functional represents various model energy functionals 
that produce physical approximations, including LDA+U and RPA. 
Similarly, we can derive 
the G-W approximation (GW) and the dynamical mean-field (DMF) approximation, 
which are discussed elsewhere. In some treatments, 
when there is a single-reference form, we have the definition of $E_{\rm EXX}$. 
Other options for the double-counting term and the exchange-correlation functional are 
i) $E_{dc}^{(1)}[\Psi] \rightarrow E_{\rm Hartree} [n_\Psi]+E_{\rm xc}[\Psi] $ keeping 
the model exchange-correlation functional effective for determining 
single-particle wave functions, and ii) 
$E_{dc}^{(1)}[\Psi] \rightarrow E_{\rm Hartree} [n_\Psi]+E_{\rm EXX}[\Psi] $ with 
$E_{\rm xc}[\Psi] \rightarrow E_{\rm EXX}[\Psi]$. The correlation part is 
explicitly expressed in the quantum fluctuation part, $E_{\rm fluc}$. 

\section{Adiabatic-connection fluctuation-dissipation-theorem with random phase approximation starting from LDA+U}

\subsection{Total energy formulas}
In this section, we consider the total energy of the Coulomb system 
when it is described by $\lambda$ integration. 
For this purpose, we analyze the expansion of the Coulomb interaction operator $\hat{V}_{\rm ee}$. 
By performing the Bloch-Wannier transformation,  
the Bloch representation in (\ref{Coulomb_expansion}) is re-expanded in the Wannier basis. 
Then, the concept of diagonal interaction as the site-diagonal is introduced. 
Assuming that the correlated band is given by the $l$th band, we have a 
separation of interaction operators into intra-band-site-diagonal contribution,  
$\hat{V}_{\rm intra-diag}^{(l)}$, intra-band-site-offdiagonal contribution, 
$\hat{V}_{\rm intra-offdiag}^{(l)}$, and inter-band contribution, 
$\hat{V}_{\rm inter-band}^{(l_1,l_2,l_3,l_4)}$. 
To obtain the last one, the band indices must satisfy  
$(l_1,l_2,l_3,l_4)\neq(l,l,l,l)$. 
We thus have the following expression:
\begin{eqnarray}
\hat{V}_{\rm ee} 
&=&\sum_l \hat{V}_{\rm intra-diag}^{(l)}
+\sum_l \hat{V}_{\rm intra-offdiag}^{(l)} \nonumber \\
&+&\sum_{(l_1,l_2,l_3,l_4)\neq(l,l,l,l)}\hat{V}_{\rm inter-band}^{(l_1,l_2,l_3,l_4)}.
\end{eqnarray}
We analyze the effect of the projection operators $P_A$ and $P_B$. 
\begin{eqnarray}
P_A \hat{V}_{\rm ee} P_A 
&=&\sum_{\epsilon_{l,{\bf k}} \le E_F} \left(\hat{V}_{\rm intra-diag}^{(l)}
+\hat{V}_{\rm intra-offdiag}^{(n)}\right) \nonumber \\
&+&\sum_{(l_1,l_2,l_3,l_4)\neq(l,l,l,l)} P_A \hat{V}_{\rm inter-band}^{(l_1,l_2,l_3,l_4)} P_A.
\end{eqnarray} 
The last contribution appears as the potential scattering of correlated 
electrons by core electrons in $S_A^{c,1}$. Similarly, we have 
\begin{equation}
P_B \hat{V}_{\rm ee} P_A = \sum_{(l_1,l_2,l_3,l_4)} P_B \hat{V}_{\rm inter-band}^{(l_1,l_2,l_3,l_4)} P_A.
\end{equation} 
We consider the Hubbard-type on-site interaction, $\hat{V}_{\rm Hubbard}$, 
its mean-field approximation, $\hat{\bar{V}}_{\rm Hubbard}$, 
and the off-diagonal contribution, $\hat{V}_{\rm Hubbard-offdiag}$. 
In our formalism of the screening, the interaction is given as follows: 
\begin{eqnarray}
\lefteqn{\hat{V}_{\rm Hubbard}} \nonumber \\ 
&=& \hat{V}_{\rm intra-diag}^{(n)} \nonumber \\
&+&\left(P_A \hat{V}_{\rm inter-band}P_B G^{2} (0^+) 
P_B\hat{V}_{\rm inter-band}P_A\right)_{\rm diag}^{(n)}. \nonumber \\
\end{eqnarray}
For the off-diagonal interaction, we have a similar expression. 
\begin{eqnarray}
\lefteqn{\hat{V}_{\rm Hubbard-offdiag}} \nonumber\\ 
&=& \hat{V}_{\rm intra-offdiag}^{(n)} \nonumber \\
&+& \left(P_A \hat{V}_{\rm inter-band}P_B G^{2} (0^+) 
P_B\hat{V}_{\rm inter-band}P_A\right)_{\rm offdiag}^{(n)}. \nonumber \\
\end{eqnarray}

Here, we consider the mean-field LDA+U model. 
\begin{eqnarray}
&&\left\{\hat{T} + \hat{V}_{\rm Hartree}[n_{\rm MF}]
+\hat{V}_{\rm xc}[n_{\rm MF}] \right. \nonumber \\ 
&&\left.
+ \hat{\bar{V}}_{\rm Hubbard} - \hat{\bar{V}}_{\rm dc}^{(3)} 
+\hat{V}_{\rm ext} \right\}|\Phi_{A}\rangle \nonumber \\
&&=E|\Phi_{A}\rangle
\end{eqnarray}
The action of $\hat{V}_{\rm ee}$ on this A state can be expanded as
\begin{eqnarray}
\lefteqn{\hat{V}_{\rm ee} |\Phi_{A}\rangle} \nonumber \\ 
&=& \sum_{\epsilon_{l,{\bf k}} \le E_F} 
\left(\hat{V}_{\rm intra-diag}^{(l)}
+\hat{V}_{\rm intra-offdiag}^{(l)}\right)|\Phi_{A}\rangle \nonumber \\
&&+
\sum_{(l_1,l_2,l_3,l_4)\neq(l,l,l,l)} P_A \hat{V}_{\rm inter-band}^{(l_1,l_2,l_3,l_4)} 
P_A |\Phi_{A}\rangle \nonumber \\
&&+
\sum_{(l_1,l_2,l_3,l_4)\neq(l,l,l,l)} P_B \hat{V}_{\rm inter-band}^{(l_1,l_2,l_3,l_4)} 
P_A|\Phi_{A}\rangle
\end{eqnarray}
Because we consider the Hubbard interaction by RPA with static approximation, 
the B state can be expressed by the two-particle Green's function:
\begin{eqnarray}
|\Phi_B\rangle 
&=& P_B G^{(2,cRPA)}(\omega=0^+) 
P_B \hat{V}_{\rm ee} |\Phi_A\rangle. 
\label{B_state_MF}
\end{eqnarray}
The normalization is given by 
\begin{equation}
|\Phi_A\rangle = C_{LDA+U}^{-1} |\tilde{\Phi}_{A}\rangle
\end{equation}

To derive the energy formula, we consider two successive $\lambda$-integration paths. 
For this purpose, we consider another $\lambda$-modified Hamiltonian. 
\begin{equation}
\hat{H}_{\lambda}^{(4)} = \lambda {\cal H}_{\rm Hubbard} + (1-\lambda) \hat{H}_1^{\rm LDA+U}. 
\label{lambda_modified_4}
\end{equation}
Here, the expectation value of the Hubbard interaction is as follows:
\begin{eqnarray}
\lefteqn{\langle \Phi_{\rm MF}| \hat{V}_{\rm Hubbard} |\Phi_{\rm MF} \rangle = \bar{E}_{\rm Hubbard}}
\nonumber \\
&=& \frac{1}{2} \langle \Phi_{\rm MF}| \hat{\bar{V}}_{\rm Hubbard} |\Phi_{\rm MF} \rangle .
\end{eqnarray}
In the double-counting term, there is a difference in the expectation value. 
\begin{eqnarray}
\lefteqn{\langle \Phi_{\rm MF}| \hat{V}_{\rm dc}^{(r,2)} |\Phi_{\rm MF} \rangle 
= \bar{E}_{\rm dc}^{(3)}} \nonumber \\
&=&
\langle \Phi_{\rm MF}| \hat{\bar{V}}_{\rm dc}^{(3)} |\Phi_{\rm MF} \rangle 
- \frac{1}{2}\sum_{i}U_{n}^{(r)} \bar{n}_{n,i}^2 .
\end{eqnarray}
Thus, we have the following relation:
\begin{eqnarray}
\lefteqn{\langle \Phi_{\rm MF}| {\cal H}_{\rm Hubbard} |\Phi_{\rm MF}\rangle} \nonumber \\
&=& E_{\rm kin}+\bar{E}_{\rm Hartree} + \bar{E}_{\rm xc} 
+ \bar{E}_{\rm Hubbard} - \bar{E}_{\rm dc}^{(3)}  + \bar{E}_{\rm ext} \nonumber \\
&=&\langle \Phi_{\rm MF}| \hat{H}_1^{\rm LDA+U} |\Phi_{\rm MF}\rangle \nonumber \\
&&- \frac{1}{2}\sum_{i}U_{n}^{(r)}\left(2\bar{n}_{n,i,\uparrow}\bar{n}_{n,i,\downarrow}
 + \bar{n}_{n,i}^2\right). 
\end{eqnarray}
In the above expression, to record the $\lambda$-integration formulas, 
we introduce the expectation values of $\hat{V}_{\rm Hartree}$, 
$\hat{V}_{\rm xc}$, and $\hat{V}_{\rm ext}$ as, 
\begin{eqnarray}
\bar{E}_{\rm Hartree} &=& \langle \Phi_{\rm MF} | \hat{V}_{\rm Hartree} | \Phi_{\rm MF} \rangle, \\ 
\bar{E}_{\rm xc} &=& \langle \Phi_{\rm MF} | \hat{V}_{\rm xc} | \Phi_{\rm MF} \rangle, and \\
\bar{E}_{\rm ext} &=& \langle \Phi_{\rm MF} | \hat{V}_{\rm ext} | \Phi_{\rm MF} \rangle, 
\end{eqnarray}
respectively.

Using (\ref{lambda_modified_2}) and (\ref{lambda_modified_4}),
the ground state energy $E_0$ of the Coulomb system is expressed as follows: 
\begin{eqnarray}
E_0 &=& \langle \Phi_{\rm MF}| \hat{H}_1^{LDA+U} |\Phi_{\rm MF}\rangle 
\nonumber \\
&+&\int_0^1 d\lambda_1 \frac{d}{d\lambda_1} 
\langle \Phi(\lambda_1) | \hat{H}_{\lambda_1}^{(4)}  |\Phi(\lambda_1)\rangle \nonumber \\
&+&\int_0^1 d\lambda_2 \frac{d}{d\lambda_2} 
\langle \Phi(\lambda_2) | \hat{\cal H}_{\lambda_2}^{(2)}  |\Phi(\lambda_2)\rangle \nonumber \\
&=&\langle \Phi_{\rm MF}| \hat{H}_1^{LDA+U} |\Phi_{\rm MF}\rangle
\nonumber \\
&+& \int_0^1 d\lambda_1 \langle \Phi(\lambda_1) |
\left( \hat{V}_{\rm Hubbard} - \hat{V}_{\rm dc}^{(r,2)} 
\right.
\nonumber \\
&&\left.
- \hat{\bar{V}}_{\rm Hubbard}^{(r)} + \hat{\bar{V}}_{\rm dc}^{(3)} \right)|\Phi(\lambda)\rangle 
\nonumber \\
&+& \int_0^1 d\lambda_1 \langle \Phi(\lambda_2) |
\left( \hat{V}_{\rm ee} - \hat{V}_{\rm Hartree} - \hat{V}_{\rm xc}
\right.
\nonumber \\
&&\left.
- \hat{V}_{\rm Hubbard} + \hat{V}_{\rm dc}^{(r,2)} \right) |\Phi(\lambda_2)\rangle. 
\label{lambda_integ_LDA+U}
\end{eqnarray}

\subsection{Review of ACFDT-RPA} 
In this subsection, we review the ACFDT-RPA method \cite{Langreth,Niquet}. The energy formula is obtained by 
letting $U$ be zero in the last section. 
\begin{eqnarray}
E_0 &=& \langle \Phi_{\rm MF}(U=0)| \hat{H}_{\rm Hubbard, U=0} |\Phi_{\rm MF}(U=0)\rangle 
\nonumber \\
&+&\int_0^1 d\lambda \frac{d}{d\lambda} 
\langle \Phi(\lambda) | \hat{\cal H}_{\lambda}^{(2)}(U_n=0)  |\Phi(\lambda)\rangle \nonumber \\
&=& 
E_{\rm kin}+\bar{E}_{\rm Hartree} + \bar{E}_{\rm xc}  + \bar{E}_{\rm ext} 
\nonumber \\
&+& \int_0^1 d\lambda \langle \Phi(\lambda) | \left( \hat{V}_{\rm ee} - \hat{V}_{\rm Hartree} 
- \hat{V}_{\rm xc}  \right)|\Phi(\lambda)\rangle 
\nonumber \\
&=&
E_{\rm kin}+E_{\rm Hartree} +E_{\rm EXX} + E_{\rm ACFDT-c}  + \bar{E}_{\rm ext}. 
\nonumber \\
\end{eqnarray}
The ACFDT correlation energy is given by 
\begin{eqnarray}
\lefteqn{E_{\rm ACFDT-c}} \nonumber \\
&=&-\frac{1}{2}\int_0^1 d\lambda \int \frac{d\omega}{2\pi} {\rm Tr} 
\left\{V_{\rm ee}\left[\chi_\lambda (i\omega)-\chi_0(i\omega)\right]\right\}.
\end{eqnarray}
Here, the $\lambda$-modified susceptibility is defined by 
\begin{eqnarray}
\lefteqn{\chi_\lambda ({\bf r},{\bf r}'; i\omega)} \nonumber \\
&=&
\chi_0 ({\bf r},{\bf r}'; i\omega) + \iint d^3rd^3r' \chi_0({\bf r},{\bf r}'; i\omega)
\nonumber \\
&&
\times K_{hxc}^\lambda ({\bf r},{\bf r}'; i\omega)\chi_\lambda({\bf r},{\bf r}'; i\omega),
\end{eqnarray}
and a kernel 
\begin{equation}
K_{hxc}^\lambda ({\bf r},{\bf r}'; i\omega)=\lambda V_{\rm ee}({\bf r},{\bf r}')
+f^\lambda_{\rm xc}[n]({\bf r},{\bf r}'; i\omega).
\end{equation}
When RPA is applied for $\chi_\lambda$ 
appearing in the evaluation of the $\lambda$ integration, 
we obtain the ACFDT-RPA correlation formula as follows:
\begin{eqnarray}
\lefteqn{E_{\rm ACFDT-RPA-c}} \nonumber \\
&=&-\frac{1}{2}\int \frac{d\omega}{2\pi} {\rm Tr} 
\left\{ \ln
\left[1-\chi_0(i\omega)V_{\rm ee}\right]+\chi_0 (i\omega)V_{\rm ee}\right\}. \nonumber \\
\end{eqnarray}

\subsection{Approximate estimation of $C_{33}$ of graphite by ACFDT-RPA+U}
With respect to graphite, by applying the treatment of the correlated band to 
the $p$ bands, we introduce several approximations. 
To identify a practical method, we simplify two $\lambda$ integrations 
appearing in (\ref{lambda_integ_LDA+U}). 
In the first integral with respect to $\lambda_1$, the mean-field Hubbard model 
of LDA+U is modified into the interacting Hubbard model. When the system remains 
a non-magnetic phase in which quasi-particle excitations are nearly described by 
the LDA+U band structure, the A state of $|\Phi(\lambda_1)\rangle$ remains 
nearly unchanged. In this situation, the following approximation is used: 
\begin{eqnarray}
\lefteqn{ \langle \Phi_{\rm MF}| \hat{H}_1^{LDA+U} |\Phi_{\rm MF}\rangle }
\nonumber \\
&+& \int_0^1 d\lambda_1 \langle \Phi(\lambda_1) |
\left( \hat{V}_{\rm Hubbard} - \hat{V}_{\rm dc}^{(r,2)} 
\right.
\nonumber \\
&&\left.
- \hat{\bar{V}}_{\rm Hubbard}^{(r)} + \hat{\bar{V}}_{\rm dc}^{(3)} \right)|\Phi(\lambda)\rangle 
\nonumber \\
&\simeq&
E_{\rm kin}+\bar{E}_{\rm Hartree} + \bar{E}_{\rm xc} 
\nonumber \\
&&+ \bar{E}_{\rm Hubbard} - \bar{E}_{\rm dc}^{(3)}  + \bar{E}_{\rm ext} . 
\end{eqnarray}
This is simply a mean-field approximation of the Hubbard interaction in 
the weak correlation regime. Then, there remains one $\lambda$ integration from 
the Hubbard model to the Coulomb model. 
\begin{eqnarray}
E_0 &\simeq& 
E_{\rm kin}+\bar{E}_{\rm Hartree} + \bar{E}_{\rm xc} + \bar{E}_{\rm Hubbard} - \bar{E}_{\rm dc}^{(3)} 
\nonumber \\
&+&\bar{E}_{\rm ext} + \int_0^1 d\lambda \langle \Phi(\lambda) | \left( \hat{V}_{\rm ee} - \hat{V}_{\rm Hartree} \right.
\nonumber \\
&&\left.
- \hat{V}_{\rm xc} - \hat{V}_{\rm Hubbard}^{(r)} 
+ \hat{V}_{\rm dc}^{(r,2)} \right)|\Phi(\lambda)\rangle 
\end{eqnarray}
We also assume that the charge density 
distribution $n({\bf r})$ given by $|\Phi(\lambda)\rangle$ is nearly unchanged along the 
$\lambda$ integration paths. 
Then, the density functional parts of the Hartree energy and the model exchange-correlation 
part are unchanged. 
\begin{eqnarray}
E_0 
&\simeq& E_{\rm kin} + \bar{E}_{\rm Hubbard} - \bar{E}_{\rm dc}^{(3)} + \bar{E}_{\rm ext}
\nonumber \\
&+&\int_0^1 d\lambda \langle \Phi(\lambda) | \left( \hat{V}_{\rm ee} 
 - \hat{V}_{\rm Hubbard}^{(r)} + \hat{V}_{\rm dc}^{(r,2)} \right)|\Phi(\lambda)\rangle 
\nonumber \\
&=& E_{\rm kin} + \bar{E}_{\rm Hubbard} - \bar{E}_{\rm dc}^{(3)} + \bar{E}_{\rm ext}
+E_{\rm r-corr}.  
\end{eqnarray}
The residual correlation energy is given by
\begin{eqnarray}
\lefteqn{E_{\rm r-corr}} \nonumber \\
&=&\int_0^1 d\lambda \langle \Phi(\lambda) | \left[ \hat{V}_{\rm ee} \right.
\nonumber \\
&-&
\left\{
\frac{1}{2} \sum_{i} U_n^{(r)} 
\left\{
:\left( 
\hat{n}_{n,i,\uparrow}+\hat{n}_{n,i,\downarrow} - \bar{n}_{n,i,\uparrow}-\bar{n}_{n,i,\downarrow}
\right)^2: 
\right. 
\right.
\nonumber \\
&& 
\left.
\left.
\left.
+ 
\left( \bar{n}_{n,i,\uparrow} + \bar{n}_{n,i,\downarrow}\right)
\right\}
\right\}
\right]|\Phi(\lambda)\rangle . 
\end{eqnarray} 

There is a practical reason to derive a simple approximation for $E_{\rm r-corr}$. 
First, a relevant scattering process for RPA correlation is produced by 
double electron-hole pair creation and the resulting bubble diagrams, 
which does not appear by a potential scattering of $\hat{V}_{\rm dc}^{(r,2)}$. 
The expectation value of this operator by $|\Phi(\lambda)\rangle$ 
should have little $\lambda$ dependence. This is because convergence in 
the charge density $n({\bf r})$ of hydrocarbon materials is known 
for comparison between DFT models and the diffusion Monte-Carlo calculation \cite{Kanai}, 
which verifies convergence in the occupation of the local orbitals $\bar{n}_{n,i,\sigma}$ 
along the $\lambda$ integration path. 
Therefore, the relevant integral kernel for the scattering process in the $\lambda$ integral 
is produced by $\hat{V}_{\rm ee}-\hat{V}_{\rm Hubbard}^{(r)}$. 
The difference between $\hat{V}_{\rm ee}$ and $\hat{V}_{\rm Hubbard}^{(r)}$ 
is given by
\begin{eqnarray}
\lefteqn{\hat{V}_{\rm ee} - \hat{V}_{\rm Hubbard}^{(r)}} \nonumber \\
&=&
f \hat{V}_{\rm intra-diag}^{(n)}
+\sum_{l\neq n} \hat{V}_{\rm intra-diag}^{(l)}
+\sum_l \hat{V}_{\rm intra-offdiag}^{(l)} \nonumber \\
&+&\sum_{(l_1,l_2,l_3,l_4)\neq(l,l,l,l)}\hat{V}_{\rm inter-band}^{(l_1,l_2,l_3,l_4)}.
\end{eqnarray}
Here, factor $f$ is approximately given by 
\begin{equation}
f \simeq 
\frac{\left(i,i \right|  V_{\rm ee}\left|i,i \right)
-\left(i,i \right|  W_{\rm cRPA}(0^+) \left|i,i \right)}
{\left(i,i \right|  V_{\rm ee}\left|i,i \right)}.
\end{equation}
In our estimation, this reduction factor is estimated to be $(7-2)/7=0.7$ 
for a $p$ orbital in graphite. 
In addition, the single-particle spectrum of the LDA+U model is 
nearly the same as the LDA spectrum. The wave functions of the $p$ bands 
are nearly unchanged by including the Hubbard terms with the double-counting correction. 
Therefore, we can use the following approximation. 
\begin{equation}
E_{\rm r-corr} \simeq E_{\rm Hartree} + E_{\rm EXX} + E_{\rm ACFDT-RPA-c}. 
\end{equation}
The exchange energy $E_{\rm EXX}$ and the ACFDT-RPA correlation energy 
$E_{\rm ACFDT-RPA-c}$ are estimated by LDA-based simulation as an approximation. 

The LDA+U+RPA total energy is given by 
\begin{eqnarray}
E_{\rm U+RPA} 
&=& E_{\rm kin} + E_{\rm Hartree} + E_{\rm EXX} + E_{\rm ACFDT-RPA-c} \nonumber \\
&&+ \bar{E}_{\rm Hubbard} - \bar{E}_{\rm dc}^{(3)} + \bar{E}_{\rm ext} \nonumber \\ 
&=& E_{\rm LDA+U}(U^{(r)}) - E_{\rm xc} \nonumber \\
&&+ E_{\rm EXX} + E_{\rm ACFDT-RPA-c} . 
\end{eqnarray} 
Here, the LDA+U energy is given by a self-consistent determination of 
$|\tilde{\Psi}_A\rangle$ with estimated $U^{(r)}$ in cRPA calculation. 
When a path of $\lambda$ integration is considered, which is reversed 
with respect to the inclusion of RPA correlation and Hubbard term correction, 
we can add the latter contribution for the ACFDT-RPA result. 
\begin{eqnarray}
E_{\rm ACFDT-RPA+U} 
&=& E_{\rm ACFDT-RPA} + \bar{E}_{\rm Hubbard} - \bar{E}_{\rm dc}^{(3)}. \nonumber \\
\end{eqnarray} 
This reversed counting of the RPA correlation and short-range correlation 
is partly rationalized by the known lack of a short-range correlation 
effect in RPA. 

\begin{figure}[tp]
\begin{center}
\includegraphics[width=85mm]{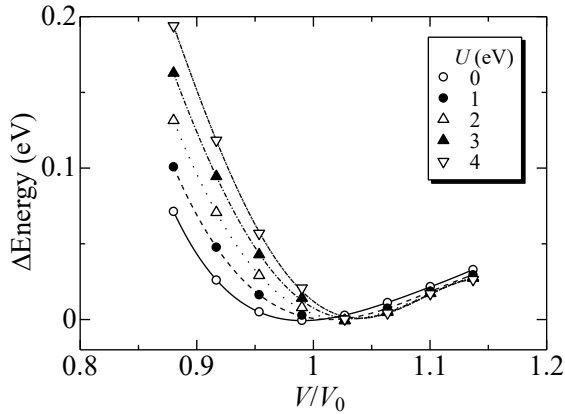}
\end{center}
\captionsetup{labelformat=empty,labelsep=none}
\caption{Fig. SB1.  Total energy curves given by the LDA+U with Wannier $p$ orbitals. Open circles, closed circles, open upward triangles, closed upward triangles, and open downward triangles represent the total energy for $U=0$, 1, 2, 3, and 4 [eV], respectively. The energy difference from the minimum value is given for each energy curve on the vertical axis. $V/V_0$ on the vertical axis is the normalized volume of the unit cell, where the measured volume of $V_0$ is 35.16 $\AA^3$.}
\label{Fig. SB1}
\end{figure}

In the evaluation of the elastic constants, we obtain the total energy for deformed 
material structures with non-zero strain (Fig.~\ref{Fig. SB1}).
When the atomic positions are changed, the 
relevant contribution in the change of the total energy can originate from 
$\bar{E}_{\rm Hubbard} - \bar{E}_{\rm dc}^{(3)}$. In fact, our formulation is given 
by the determination of local operators in the material. We determine 
the Kohn-Sham wave functions as the band structure, and the Wannier transformation 
is used to introduce the U terms \cite{Cococcioni}. 
Structural modification causes a significant change of 
the wave functions. Therefore, $\bar{E}_{\rm Hubbard} - \bar{E}_{\rm dc}^{(3)}$ creates 
a relevant change in the result of $C_{33}$. 
This contribution nearly disappears when the atomic wave function determined 
prior to the bulk calculation is used to introduce the U terms. 
A short-range correlation appears with the static modification of the $\pi$ orbitals 
with an adiabatic change of the graphite structure. This 
approximation is generally justified for the practical conditions 
appearing in the ultrasonic experiment. 

In the simulation, we adopt the Quantum ESPRESSO package \cite{QE} for the LDA+U calculation, 
VASP \cite{vasp_1,vasp_2,vasp_3} for the ACFDT-RPA calculation, and 
RESPACK \cite{respack1,respack2,respack3,respack4,respack5} for the constrained RPA calculation. 

